\documentclass[final]{article}
\usepackage{amsthm,amsmath,amsfonts}

\newtheorem*{defn}{Definition}

\newtheorem*{alg}{Algorithm}\newtheorem*{lem}{Lemma}
\newtheorem*{Main}{Main Theorem}
\newtheorem*{cor}{Corollary}
\newcommand{\ket}[1]{| #1 \rangle}
\newcommand{\bra}[1]{\langle #1 |}

\title{\bf{A Scheme of Cartan Decomposition for $su(N)$}}
\author{ Zheng-Yao Su \thanks{National Center for High-Performance Computing, Hsinchu, Taiwan, R.O.C.} \thanks{Department of Physics, National Tsing Hua University, Hsinchu, Taiwan, R.O.C.}
         \thanks{Email: zsu@nchc.gov.tw} }\date{}
\begin{document}

\maketitle

\begin{abstract}
 \hspace{0.1cm}A scheme to perform the Cartan decomposition for
 the Lie algebra $su(N)$ of arbitrary finite dimensions
 is introduced. The scheme is based on two algebraic structures,
 the {\em conjugate partition} and the {\it quotient algebra},
 that are easily generated by a Cartan subalgebra and
 generally exist in $su(N)$.
 In particular, the Lie algebras $su(2^p)$ and every $su(2^{p-1}<N<2^p)$
 share the isomorphic
 structure of the quotient algebra.
 This structure enables
 an efficient algorithm for the recursive and exhaustive construction of Cartan decompositions.
 Further with the scheme,
 a unitary transformation in $SU(N)$ can be recursively
 decomposed into
 a product of certain designated operators, {\em e.g.}, {\it local} and {\it nonlocal} gates.
 Such a recursive decomposition of a transformation implies an evolution {\em path} on the manifold of the group.
\end{abstract}

\pagestyle{myheadings} \thispagestyle{plain} \markboth{ Zheng-Yao Su, and  Ming-Chung Tsai}{A Scheme of Cartan Decomposition}

\section{Introduction}
\renewcommand{\theequation}{\arabic{section}.\arabic{equation}}
\setcounter{equation}{0}
 The evolution of a closed quantum system of finite dimensions is governed
 by a unitary transformation in the Lie group $SU(N)$.
 Portraying the evolution {\em path} on the manifold of the group, normally with very high
 dimensions, will be of great help
 to the investigation of the system.
 Geometrically, the set of generators for the Lie algebra $su(N)$
 serves as an intrinsic {\em moving frame} on the manifold of the group $SU(N)$ \cite{Spivak}.
 Besides offering a geometric basis, algebraically   
 $su(N)$ admits the Cartan decomposition that has fundamental importance in
 the treatment of Lie algebras and Lie groups \cite{Helgason,Knapp}.
 The geometric concepts and algebraic skills of the Cartan decomposition have recently been
 brought into the areas of quantum control and quantum information \cite{Khaneja2}.
 Much attention and effort are thus dedicated to a wider scope of applications
 \cite{Khaneja1,Khaneja3,Zhang}.
 Furthermore, a large programme for quantum computation has very
 recently been proposed that is mainly depicted in terms of the geometric language
 \cite{Nielsen1,Nielsen2}. An effective device, in a geometric or an algebraic
 framework, for the exploration on the immense manifold of unitary transformations
 has become desirable.

 In this article, a general and simple scheme of the Cartan decomposition is introduced.
 This scheme is applicable to the Lie algebra $su(N)$ of arbitrary dimensions without
 being restricted to a power of $2$.
 In the scheme, an arbitrary Cartan
 subalgebra in $su(N)$ can create
 a particular form of partition of the algebra, the conjugate partition, that is
 composed of a certain number of abelian subalgebras.
 These abelian subalgebras further form an algebraic structure,
 the quotient algebra, due to the closure under the operation of the Lie bracket.
 Importantly, every Lie algebra $su(2^{p-1}<N<2^p)$ shares with $su(2^p)$
 the isomorphic structure of the quotient algebra.
 This structure makes straightforward and systematic the exhaustive construction of Cartan decompositions in $su(N)$.
 When the decomposition is recursively applied to the final stage,
 only an abelian subalgebra remains.
 It immediately leads to a recursive decomposition of the Lie
 group $SU(N)$, such that a unitary transformation is factorized into a product
 of actions respectively contributed only by a generator of $su(N)$.
 A recursive decomposition of a transformation implies a {\em path} on the manifold of the
 Lie group $SU(N)$, which is adjustable by the choice of the abelian subalgebra
 designated at each level of the decomposition.
 The scheme may have provided a tool for the journey on the manifold of $SU(N)$.

\section{Conjugate Partition}\renewcommand{\theequation}{\arabic{section}.\arabic{equation}}
\setcounter{equation}{0}\hspace{0.3cm}
 To have a better access to the scheme, the exposition will begin
 with the examples of decompositions for $su(6)$ and $su(8)$.
 Here, $su(8)$ is the Lie algebra of the transformation group $SU(8)$ acting on a $3$-qubit
 system, and $su(6)$ is the algebra of the group $SU(6)$ acting on a system composed of
 a $2$-state and a $3$-state quantum particles. Although the scheme is representation
 independent, simply for convenience the exposition will be given in the spinor representation.
 The generators of these algebras are hence written in tensor products
 of the Pauli and Gell-Mann matrices.

 The first step is to take a {\bf \em center subalgebra}, denoted as ${\cal A}$,
 which is a maximal abelian subalgebra,
 a Cartan subalgebra, arbitrarily chosen from
 the algebra to be decomposed.
 With this choice of the center subalgebra, more abelian subalgebras will be generated.
 An abelian subalgebra of $7$ elements, listed in the central column of Fig.\ref{stepsu8},
 is an example of the center subalgebra for $su(8)$,
 ${\cal A}=\{\sigma_{3}\otimes I\otimes I, I\otimes\sigma_{3}\otimes I,
            I\otimes I\otimes\sigma_{3}, \sigma_{3}\otimes\sigma_{3}\otimes I,
            \sigma_{3}\otimes I\otimes \sigma_{3}, I\otimes\sigma_{3}\otimes\sigma_{3},
             \sigma_{3}\otimes\sigma_{3}\otimes\sigma_{3}\}$.
 It is easy to construct an abelian subalgebra
 for a Lie algebra, actually a maximal one, by collecting
 all its diagonal generators. The second step is to, outside the center subalgebra,
 take an arbitrary generator as a
 {\bf {\em seed}}, for example $ I\otimes I\otimes \sigma_{1}$ at the left hand side
 of the center subalgebra in Fig.\ref{stepsu8}a. Four other generators are produced, listed
 at the right hand side, by calculating the commutator of the seed with each element
 in the center subalgebra, {\em i.e.,}
 mapping the seed to another vector space by the adjoint representation $ad_{\cal A}$,
 $[I\otimes I\otimes \sigma_{1}, \sigma_{3}\otimes I \otimes I] =
       -2iI\otimes I\otimes \sigma_{2}$,
       $[I\otimes I\otimes \sigma_{1}, \sigma_{3}\otimes \sigma_{3} \otimes I] =
       -2i \sigma_{3}\otimes I \otimes\sigma_{2} $, {\em etc}.
 Then, in the reversing step,
 $3$ more generators are added to the LHS column as shown in Fig.\ref{stepsu8}b after
 taking any one of the $4$ generators in the RHS column and calculating the commutators
 of this generator with those in the center subalgebra.

 The first good property is that the elements in these two new columns respectively
 form an abelian subalgebra.
 The $4$ generators of such an abelian subalgebra also build a vector space.
 Let $W_1$ and $\hat{W}_1$ denote the vector spaces respectively spanned
 by the subalgebras in the LHS and RHS columns; each of this pair is considered
 the {{\bf \em conjugate}} by the other, noting that $[W_1,\hat{W}_1]\subset \cal{A}$.
 Following the same procedure, more conjugate pairs of abelian subalgebras
 $\{W_i,\hat{W}_i\}$
 are produced by taking a seed outside the center subalgebra
 and the existing conjugate pairs, from $\{W_{1},\hat{W}_{1}\}$
 to $\{W_{i-1},\hat{W}_{i-1}\}$, and calculating the required  commutators.
 The continued exposition will temporarily be based on the current version of
 the {\bf \em algorithm primitive}.

 This procedure leads to a partition
 of the original algebra as given in Fig.\ref{csu8}.
 Stated in the following Lemma,
 such a {\bf \em conjugate partition} always exists for
 the Lie algebra $su(N)$, which
 will be proved with the Main Theorem.

 \begin{lem}\label{lemma}
 With an arbitrary maximal abelian subalgebra ${\cal A}$ taken as
 the {\bf\em center subalgebra}, the Lie algebra $su(N)$ has
 a {\bf\em conjugate partition} consisting of ${\cal A}$
 and a finite number, $q$,
 of {\bf\em conjugate pairs} of abelian subalgebras
 $\{W_i,\hat{W}_i\}, 1\leq i\leq q$, i.e.,
 \begin{equation}\label{L1}
 su(N)= {\cal A} \oplus W_{1} \oplus \hat{W}_{1} \oplus\cdots\oplus W_{i} \oplus \hat{W}_{i}  \oplus
   \cdots \oplus W_{q} \oplus \hat{W}_{q}
 \end{equation}
 and no intersection between any two of the subalgebras, where
 the abelian subalgebras respectively form a vector space and follow the
 conditions, $\forall\hspace{0.1cm} 1\leq i\leq q$ and
 no order for entries in the commutator,
 \begin{equation}\label{L2}
 [W_{i},{\cal A}]\subset \hat{W}_{i}, \ \ [\hat{W}_{i},{\cal A}]\subset W_{i}, \ \
 {\rm and} \ \ [W_{i},\hat{W}_{i}]\subset {\cal A}.
 \end{equation}
 \end{lem}

 \section{Quotient Algebra}
 \renewcommand{\theequation}{\arabic{section}.\arabic{equation}}
 \setcounter{equation}{0}\hspace{0.3cm}
 In addition to the partition composed of conjugate pairs of abelian subalgebras,
 of great interest is an algebraic structure embedded in these subalgebras.
 An easy check is that, for the partition listed
 in Fig.\ref{csu8}, the commutator of any elements from two different subalgebras
 must be either $\{0\}$ or in a third subalgebra.
 It reveals the existence of an algebraic structure, the {\bf {\em quotient algebra}},
 based on which
 constructing Cartan decompositions becomes straightforward.

\begin{figure}
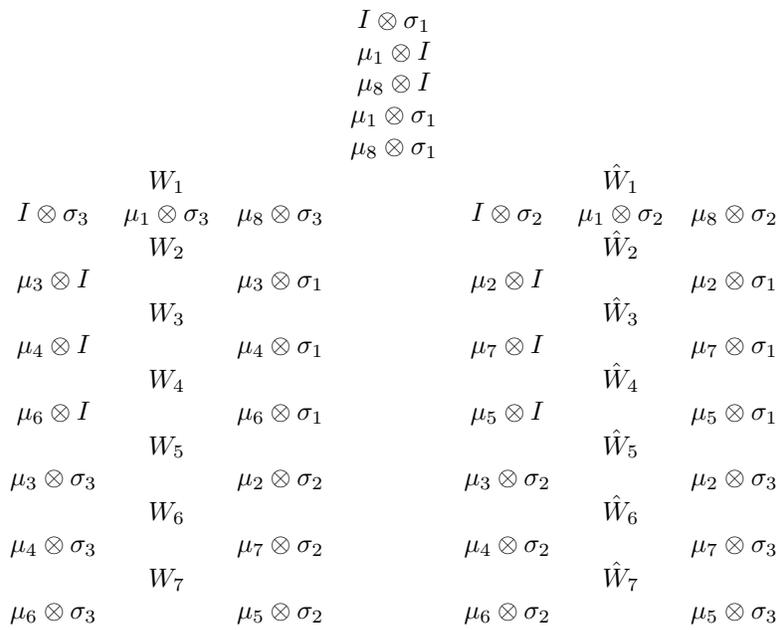

\begin{center}
\[\begin{array}{ccccccc}
 & & &  I\otimes \sigma_{1}  & & &\\
 & & &   \mu_{1} \otimes I   & & & \\
 & & &   \mu_{8} \otimes I  & & & \\
 & & & \mu_{1} \otimes \sigma_{1} & & & \\
& & & \mu_{8} \otimes \sigma_{1} & & & \\

& W_{1} & & & & \hat{W}_{1} & \\
 I \otimes \sigma_{3}  &  \mu_{1}\otimes \sigma_{3}  &  \mu_{8}\otimes \sigma_{3}  & &  I\otimes \sigma_{2}  &  \mu_{1}\otimes \sigma_{2}  &  \mu_{8} \otimes \sigma_{2} \\
& W_{2} & & & & \hat{W}_{2} & \\
  \mu_{3} \otimes I  & &  \mu_{3}\otimes\sigma_{1} & &  \mu_{2} \otimes I  & & \mu_{2} \otimes \sigma_{1}\\
& W_{3} & & & & \hat{W}_{3} & \\
  \mu_{4} \otimes I  & & \mu_{4}\otimes \sigma_{1} & &   \mu_{7} \otimes I & & \mu_{7} \otimes \sigma_{1}\\
& W_{4} & & & & \hat{W}_{4} & \\
  \mu_{6} \otimes I  & & \mu_{6}\otimes \sigma_{1} & &   \mu_{5} \otimes I  & & \mu_{5}\otimes \sigma_{1}\\
& W_{5} & & & & \hat{W}_{5} & \\
   \mu_{3} \otimes \sigma_{3} & & \mu_{2}\otimes \sigma_{2} & &  \mu_{3}  \otimes \sigma_{2}  & & \mu_{2}\otimes \sigma_{3}\\
& W_{6} & & & & \hat{W}_{6} & \\
  \mu_{4} \otimes \sigma_{3}  & & \mu_{7}\otimes \sigma_{2} & &   \mu_{4} \otimes \sigma_{2}  & & \mu_{7}\otimes \sigma_{3}\\
& W_{7} & & & & \hat{W}_{7} & \\
  \mu_{6} \otimes  \sigma_{3} & & \mu_{5}\otimes \sigma_{2} &
&   \mu_{6}\otimes  \sigma_{2} & & \mu_{5}\otimes \sigma_{3}
\end{array} \]
\caption{A conjugate partition and a quotient algebra given
 by a non-diagonal center subalgebra of $su(6)$.}\label{ncsu6}
\end{center}
\end{figure}

 \begin{defn}\label{defn}
 For a conjugate partition given by the center subalgebra ${\cal A}$
 and with $q$ conjugate pairs
 of abelian subalgebras $\{W_i,\hat{W}_i\}, 1\leq i\leq q$, the subalgebras
 construct a {\bf \em quotient algebra},
 denoted as a multiplet of partition
 $\{{\cal Q(A};q)\}\equiv\{{\cal A}; W_i,\hat{W}_i,1\leq i\leq q\}$, if
  the {\bf\em condition of closure} under the operation of the commutator is satisfied:
 $\forall\hspace{0.1cm} 1\leq i,j\leq q, \hspace{0.2cm}
  \exists\hspace{0.1cm} 1\leq k \leq q$,  s.t.
 \begin{equation}\label{D1}
 [W_i,W_j]\subset \hat{W}_k, \
 [W_i,\hat{W}_j]\subset W_k, \ \ {\rm and} \ \
 [\hat{W}_i,\hat{W}_j]\subset \hat{W}_k;
 \end{equation}
 here the center subalgebra ${\cal A}$ acts as the zero element of the operation
 within a conjugate pair,
 i.e.,
 $\forall\hspace{0.1cm} 1\leq i\leq q,$
\begin{equation}\label{D2}
 [W_{i},{\cal A}]\subset \hat{W}_{i}, \ \ [\hat{W}_{i},{\cal A}]\subset W_{i}, \ \
 {\rm and} \ \ [W_{i},\hat{W}_{i}]\subset {\cal A}.
\end{equation}
 \end{defn}
 In the following the notation 
 will be abbreviated
 as $\{{\cal Q(A)}\}$; the former form is resumed only when the number
 of conjugate pairs should be noted.
 This structure exists not only in $su(N=2^p)$ but also in $su(N\neq 2^p)$.
 Similarly, the center subalgebra consisting of $5$ diagonal generators of $su(6)$,
 $\{I\otimes \sigma_3, \mu_3\otimes I, \mu_8\otimes I,\mu_3\otimes\sigma_3,
    \mu_8\otimes\sigma_3\}$, creates a partition as shown in Fig.\ref{csu6}, where
 a quotient algebra resides.
 The generators $\mu_j, j=1,2,\dots,8$, for $su(3)$ denote the Gell-Mann matrices,
 ref. Appendix A.
 Here the vector spaces
 $W_i$ or $\hat{W}_i$ are spanned by $2$ or $3$ generators, rather than equally
 $4$ as those for $su(8)$. To construct a conjugate partition and the corresponding
 quotient algebra, the center subalgebra is not restricted to only diagonal generators.
 Figs.\ref{ncsu8} and \ref{ncsu6} demonstrate the conjugate partitions and quotient algebras
 for $su(8)$ and $su(6)$ that are made by taking
 other choices of center subalgebras. This observation foretells
 the Main Theorem.

 \begin{Main}\label{main}
 Every Lie algebra $su(N)$ has the structure of the quotient algebra,
 and every maximal abelian subalgebra ${\cal A}\subset su(N)$ can generate
 a quotient algebra $\{{\cal Q(A)}\}$.
 \end{Main}
 \begin{proof}
 The Main Theorem and the Lemma will be proved first for the Lie algebra $su(N=2^p)$ and
 then extended to the algebra $su(N\neq 2^p)$ as the Corollary.

 For every maximal abelian subalgebra ${\cal A}\in su(N)$, there exists a transformation in $SU(N)$ that maps ${\cal A}$ to
 the {\bf \em intrinsic coordinate} or the {\bf \em eigenspace} of ${\cal
 A}$, where all elements of the subalgebra ${\cal A}$ are simultaneously
 diagonalized. While every maximal abelian subalgebra has its own  intrinsic coordinate which can be connected to that of any other maximal abelian subalgebra by a unitary
 transformation. Equivalently, once a coordinate is decided, the center subalgebra consisting of only diagonal operators
 is regarded as the {\bf \em intrinsic center subalgebra} and has a reserved notation ${\cal
 C}$. Although coordinate independent, the structure and properties of the quotient algebra are better discerned in the eigenspace of the chosen center
 subalgebra. The exposition thus starts in this coordinate.

 \paragraph{$\lambda$-Representation.}
 The vector space of the intrinsic center subalgebra ${\cal C}$ of
 $su(4)$ for example is
 spanned by the $3$ diagonal operators,
 $I\otimes\sigma_3=$ $diag\{1,-1,$ $1,-1\}$,
    $\sigma_3\otimes\sigma_3 =$ $ diag\{1,-1,$ $-1,1\}$  and
    $\sigma_3\otimes I =$ $ diag\{1,1,-1,-1\}$,
 or equivalently by another set of $3$ independent operators
    $diag\{1,-1,0, 0\}$,
    $diag\{1, 0,-1,$ $0\}$  and  $diag\{1,0,0,-1\}$.
 In general, the intrinsic center subalgebra ${\cal C}$ of the Lie algebra $su(N)$
 is spanned
 by the $N-1$, $N\times N$, diagonal operators: 
 $diag\{1,-1,$ $0,\cdots,0\}$, $diag\{1,0,-1,0,\cdots,0\}$,
 $\dots$ and $diag\{1,0,\cdots,0,-1\}$.
 According to the algorithm primitive, calculating the commutators of a
 seed generator with ${\cal C}$,
 or mapping the seed by the adjoint representation $ad_{\cal C}$, produces
 the generators in the conjugate vector space.
 To examine this operation, 
  the so-called
 {\bf $\lambda$-{\em representation}}, an extension of the well-known Gell-Mann matrices,
 is employed, referring to Appendix A for more details.
 In this representation, a $\lambda$-generator $\lambda_{ij}$,
 an off-diagonal $N\times N$ matrix,
 plays the role of $\sigma_1$, whose
 the only two nonzero entries, the $(i,j)$-th and $(j,i)$-th, are both assigned $1$.
 Another $\lambda$-generator $\hat{\lambda}_{ij}$, the {\em conjugate}
 of $\lambda_{ij}$ and an off-diagonal $N\times N$ matrix too,
 takes the role of $\sigma_2$ and has nonzero values only at the
 $(i,j)$-th and the $(j,i)$-th entries, respectively assigned $-i$ and $i$.
 According to eqs.\ref{A1}-\ref{A3},
 the adjoint representation $ad_{\cal C}$ leaves a $\lambda_{ij}$
 or a $\hat{\lambda}_{ij}$ {\em invariant within the pair}.
 The $\lambda$-generators may well be acting as building blocks to 
 form the conjugate pairs of abelian subalgebras.
 In terms of these generators, some {\em combinatorial traits} of the scheme
 will be read in the following construction of
 conjugate partitions and quotient algebras.

 \paragraph{Binary Partitioning.}
 A set of the $\lambda$-generators by an appropriate grouping suffices to serve as a
 basis for the vector space of an abelian subalgebra. Such a grouping
 implies a corresponding partition on the subscripts of the $\lambda$-generators.
 Since two of these generators commute as long as no repetition in the subscripts,
 a maximal abelian subalgebra in $su(N)-{\cal C}$ is made by taking
 any one set of $N/2$ elements \{$\lambda_{ij}$ or $\hat{\lambda}_{kl}$: the subscripts
 of the generators are given by a partition of the integers
 from $1$ to $N$, $i,j,k,l=1,2,\dots,N$\}.
 For instance, two sets of $4$ generators in $su(8)$,
 $\{\lambda_{15},\hat{\lambda}_{26},\lambda_{37},\hat{\lambda}_{48}\}$ and
 $\{\lambda_{16},\hat{\lambda}_{28},\hat{\lambda}_{37},\hat{\lambda}_{45}\}$,
 respectively build an abelian subalgebra.
 Although both partitions of the subscripts lead to the creation of an abelian subalgebra,
 only the former kind fits the further use, to support quotient algebras.
 Generators of conventional types bring the clue.
 Take one spinor generator $\sigma_3\otimes I\otimes\sigma_1 \in su(8)$ for example,
 which reads as $\lambda_{12}+\lambda_{34}-\lambda_{56}-\lambda_{78}$
 in the $\lambda$-representation and is a vector in the space $W_1$ spanned
 by the $4$ generators $\lambda_{12},\lambda_{34},\lambda_{56},$ and $\lambda_{78}$.
 By the algorithm primitive, the $4$ independent generators for
 the conjugate vector space $\hat{W}_1$
 are produced from the commutator $[\sigma_3\otimes I\otimes\sigma_1,{\cal C}]$.
 For owing to eqs.\ref{A1} and \ref{A2}, the required $+/-$ parities assigned to $\lambda_{ij}$s
 are provided by the mapping $ad_{\cal C}$    
 when going through all the, $N-1$, independent generators of ${\cal C}$.
 Conversely, the commutator $[\hat{g},{\cal C}]$,
 $\hat{g}$ being any one of the $4$ conjugate generators in $\hat{W}_1$,
 adds another $3$ independent generators to
 the vector space $W_1$.
 It is critical to note that the subscripts of
 the $\lambda$-generators so produced in the same conjugate pair share
 a common binary pattern of partition,
 here particularly termed as the {\bf \em binary partitioning}.
 For the vector space $W_1$, the subscripts of the $\lambda$-generators
 have the common pattern of {\em bit-wise addition} $i'+001=j'$,
 where $i'=i-1$ and $j'=j-1$ are written in their binary expressions.
 Since the mapping $ad_{\cal C}$ leaves
 a $\lambda$-generator invariant in its conjugate pair,
 the identical subscript pattern of binary partitioning shall be
 retained by all the generators belonging to the same pair.
 In other words, {\em the binary partitioning is an invariant under
 the adjoint representation} $ad_{\cal C}$.
 It is legitimate to {\em encode} such a subscript pattern by a binary string and
 attach the string to the conjugate pair.
 The vector spaces $W_1$ and $\hat{W}_1$ are therefore redenoted as
 as $W_{001}$ and $\hat{W}_{001}$.

\paragraph{Conjugate Partition.}
 The continued step is to pick a spinor generator as a seed
 outside this conjugate pair and the center subalgebra,
 for example $\sigma_3\otimes\sigma_1\otimes\sigma_1=
          \lambda_{14}+\lambda_{23}-\lambda_{58}-\lambda_{67}$ where the
 subscripts of the $\lambda$-generators keep another pattern of binary partitioning
 $i'+011=j'$, with $i'=i-1$ and $j'=j-1$ in
 their binary expressions. Then the conjugate pair
 $W_{011}$ and $\hat{W}_{011}$ are formed
 by operations of the commutators required in the algorithm primitive.
 Accordingly, by this procedure all the
 vector spaces of abelian subalgebras,
 associated to the strings from $001$ to $111$, appear pair by pair, ref. Fig.\ref{Gsu8}.
 These $7$ conjugate pairs of vector spaces along with the center subalgebra ${\cal C}$
 build a conjugate partition in $su(8)$.
 This procedure is generally applicable to the Lie algebra $su(2^p)$, whose
 any one spinor generator in the $\lambda$-representation
 specifies a subscript pattern of binary partitioning.
 Let a binary string of $p$ digits encode the subscript pattern.
 Say associated to the string $\zeta$,
 an arbitrary spinor generator $g_\zeta\in su(2^p)$ is taken as a seed
 outside ${\cal C}$ and the existing conjugate pairs.
 With the application of eqs.\ref{A1} and \ref{A2},
 a new conjugate pair $\{W_{\zeta},\hat{W}_{\zeta}\}$ is formed
 by the mappings $[g_\zeta,{\cal C}]$ and $[\hat{g}_\zeta,{\cal C}]$,
 where $\hat{g}_\zeta$ is a spinor generator produced from the former commutator.
 Such a step is recursively applied until no generator is left and hence
 all the $2^p-1$  strings of binary partitioning
 and the $2^p-1$ associated conjugate pairs have appeared.
 A conjugate partition therefore completes in the Lie algebra $su(2^p)$, eq.\ref{L1}.
 The property of {\em the invariance within a conjugate pair}, eq.\ref{L2},   
 is derived from eqs.\ref{A1}-\ref{A3}.

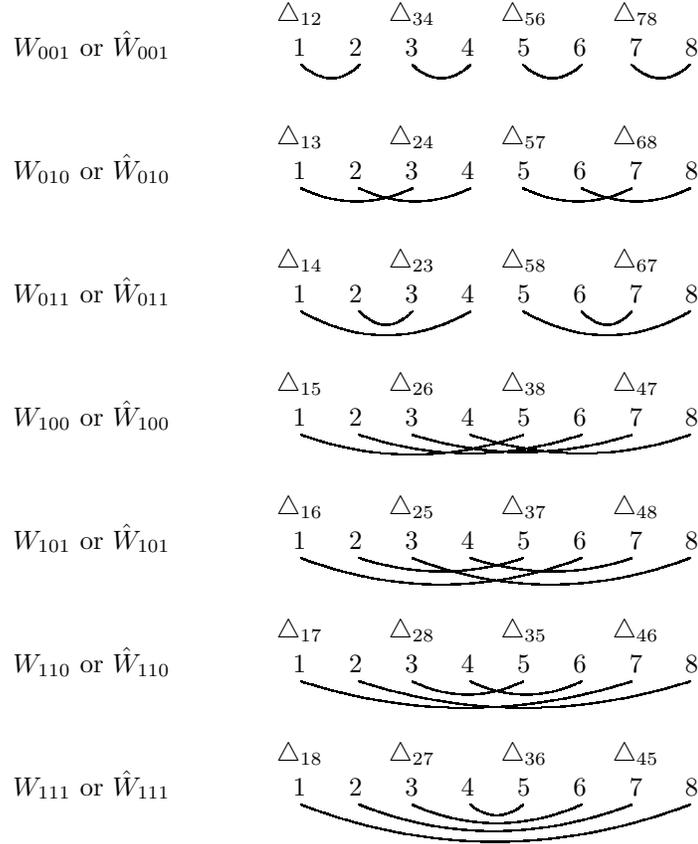
\begin{figure}
\begin{center}
\[\begin{array}{cccccccccc}
& & \triangle_{12} & & \triangle_{34} & & \triangle_{56} & &
\triangle_{78}&\\ W_{001} \text{ or } \hat{W}_{001}& & 1 & 2 & 3 &
4 & 5 & 6 & 7 & 8\\
\begin{picture}(100,30)(34,0)
\qbezier(163,30)(175,20)(185,30)\qbezier(205,30)(216,20)(227,30)\qbezier(247,30)(258,20)(269,30)\qbezier(288,30)(299,20)(310,30)
\end{picture}
& & \triangle_{13} & & \triangle_{24} & & \triangle_{57} & &
\triangle_{68}&\\ W_{010} \text{ or } \hat{W}_{010} & & 1 & 2 & 3
& 4 & 5 & 6 & 7 & 8\\
\begin{picture}(100,30)(34,0)
\qbezier(163,30)(184,20)(205,30)\qbezier(185,30)(206,20)(227,30)\qbezier(247,30)(268,20)(288,30)\qbezier(269,30)(290,20)(310,30)
\end{picture}
& & \triangle_{14} & & \triangle_{23} & & \triangle_{58} & &
\triangle_{67}&\\ W_{011} \text{ or } \hat{W}_{011} & & 1 & 2 & 3
& 4 & 5 & 6 & 7 & 8\\
\begin{picture}(100,30)(34,0)
\qbezier(163,30)(195,12)(227,30)\qbezier(185,30)(195,20)(205,30)\qbezier(247,30)(279,12)(310,30)\qbezier(269,30)(279,20)(288,30)
\end{picture}
& & \triangle_{15} & & \triangle_{26} & & \triangle_{38} & &
\triangle_{47}&\\ W_{100} \text{ or } \hat{W}_{100} & & 1 & 2 & 3
& 4 & 5 & 6 & 7 & 8\\
\begin{picture}(100,30)(34,0)
\qbezier(163,30)(205,15)(247,30)\qbezier(185,30)(227,16)(269,30)\qbezier(205,30)(247,17)(288,30)\qbezier(227,30)(270,16)(310,30)
\end{picture}
& & \triangle_{16} & & \triangle_{25} & & \triangle_{37} & &
\triangle_{48}&\\ W_{101} \text{ or } \hat{W}_{101} & & 1 & 2 & 3
& 4 & 5 & 6 & 7 & 8\\
\begin{picture}(100,30)(34,0)
\qbezier(163,30)(216,10)(269,30)\qbezier(185,30)(216,20)(247,30)\qbezier(205,30)(257,10)(310,30)\qbezier(227,30)(259,20)(288,30)
\end{picture}
& & \triangle_{17} & & \triangle_{28} & & \triangle_{35} & &
\triangle_{46}&\\ W_{110} \text{ or } \hat{W}_{110} & & 1 & 2 & 3
& 4 & 5 & 6 & 7 & 8 \\
\begin{picture}(100,30)(34,0)
\qbezier(163,30)(225,10)(288,30)\qbezier(185,30)(247,10)(310,30)\qbezier(205,30)(226,20)(247,30)\qbezier(227,30)(248,20)(269,30)
\end{picture}
& & \triangle_{18} & & \triangle_{27} & & \triangle_{36} & &
\triangle_{45}&\\ W_{111} \text{ or } \hat{W}_{111} & & 1 & 2 & 3
& 4 & 5 & 6 & 7 & 8
\\
\begin{picture}(100,30)(34,0)
\qbezier(163,30)(236,2)(310,30)\qbezier(185,30)(235,10)(288,30)\qbezier(205,30)(237,16)(269,30)\qbezier(227,30)(237,22)(247,30)
\end{picture}
\end{array}\]\\
\caption{Grouping the $\lambda$-generators of $su(8)$ into $7$
conjugate pairs
 according to the binary partitioning,
 the $\triangle$ here being a generator, either a $\lambda_{ij}$ or a $\hat{\lambda}_{ij}$;
 the commutator of any two conjugate pairs, associated to the $3$-digit strings
 $\zeta$ and $\eta$, falling in a $3$rd pair, associated to
 the bitwise-addition string  $\zeta +\eta$,
 by the condition of closure.}\label{binarysu8}
\end{center}
\end{figure}

\paragraph{Quotient Algebra.}
 Most importantly, the binary partitioning
 encodes the structure of the quotient algebra.
 The structure becomes clear as being  {\em translated}
 into the language of binary partitioning.
 For three vector spaces of abelian subalgebras
 associated to  three $p$-digit strings of binary partitioning
 $\zeta,\eta$ and $\xi$,
 the condition eq.\ref{D1} is rewritten as
 $[W_{\zeta},W_{\eta}]\subset\hat{W}_{\xi}$,
 $[W_{\zeta},\hat{W}_{\eta}]\subset W_{\xi}$, and
 $[\hat{W}_{\zeta},\hat{W}_{\eta}]\subset\hat{W}_{\xi}$.
 According to eqs.\ref{A4}-\ref{A6}, the commutator of two
 $\lambda$-generators in $su(2^p)$
 with the subscripts $(i,j)$ and $(j,k)$, or $(k,j)$, yields
 a $\lambda$-generator with the subscript $(i,k)$; here
 $i,j$ and $k$ are three distinct integers.
 Suppose that the subscript $(i,j)$ is associated to the string
 $\zeta$ and $(j,k)$ to $\eta$, and thus there exist the
 relations of bit-wise additions, $i'+j'=\zeta$ and $j'+k'=\eta$,
 $i'=i-1,j'=j-1$ and $k'=k-1$ in their binary expressions.
 The string attached to the yielded subscript $(i,k)$
 is then coerced to be $\zeta +\eta$, for $j'+j'=0$.
 Accordingly, characterizing the {\bf \em closure} of the
 abelian-subalgebra vector spaces under the operation of the commutator,
 the condition is satisfied whenever the relation of
 bit-wise addition  $\zeta + \eta = \xi$ holds. 
 To be short, the binary partitioning encodes
 the condition of closure, eq.\ref{D1}, as
 \begin{equation}\label{D15}
 [W_{\zeta},W_{\eta}]\subset\hat{W}_{\zeta +\eta}, \
 [W_{\zeta},\hat{W}_{\eta}]\subset W_{\zeta +\eta}, \ \ {\rm and}\ \
 [\hat{W}_{\zeta},\hat{W}_{\eta}]\subset\hat{W}_{\zeta +\eta}.
 \end{equation}
 The condition exhibits a property of binary combinatorics in quotient algebras.
 An example for $su(8)$ is illustrated in Fig.\ref{binarysu8}.

 That the center subalgebra ${\cal C}$ is the {\em zero element}
 within a conjugate pair under the operation
 of the commutator, eq.\ref{D2} repeats eq.\ref{L2} of the
 conjugate partition.
 Therefore, the quotient algebra ${\cal \{Q(C};2^p-1)\}$ given by, in fact ``partitioned by,"
 ${\cal C}$ is constructed. Any other quotient algebra ${\cal \{Q(A)\}}$ given
 by a maximal abelian subalgebra ${\cal A}$ is connected to the former
 by a unitary transformation $U\in SU(N)$,
 namely the equivalence via a conjugation mapping,
 ${\cal \{Q(A)\}}=U^{\dagger}{\cal \{Q(C)\}}U$
 with $U{\cal A}U^{\dagger}={\cal C}$.

 \paragraph{Subscript Multiplication.}
 Apparently there are many more options of conjugate partitions.
 To fulfill the properties in the Lemma, the binary partitioning is
 not a unique way to arrange the subscripts.
 As aforementioned, a conjugate partition is created as long as
 the pairs of the $\lambda$-generators, $\lambda_{ij}$ and $\hat{\lambda}_{ij}$,
 in every
 conjugate pair of vector spaces have no repetition in their subscripts.
 However, to further embed the structure of quotient algebras,
 the consistence in the subscripts, {\em i.e.}, following the binary partitioning
 or its equivalence, is necessary.
 An abelian-subalgebra vector space $V\subset su(8)$ for example is spanned
 by $4$ generators,
 $V=span\{\lambda_{16},\lambda_{25},\lambda_{37}.\lambda_{48}\}\subset W_{100}\cup W_{101}$.
 This grouping of subscripts is considered inconsistent owing to bearing
 two different strings of binary partitioning.
 As a result, the commutator $[V,W_{011}]$ yields $8$ generators that a half of them
 are in the space $W_{110}$ and another half in $W_{111}$.
 This inconsistence causes the partition to violate the condition of closure.
 It is shown in the following that, to accommodate
 the structure of the quotient algebra given by ${\cal C}$,
 the subscripts of the $\lambda$-generators are obliged to follow the
 binary partitioning or its permutations.

 A worthy reexamination is that how the condition of closure is met
 under the operation of the commutator for any two abelian subalgebras
 in the conjugate partition given by an intrinsic center subalgebra.
 Due to eqs.\ref{A4}-\ref{A6}, the operation can
 be reduced to as simple as a {\em subscript multiplication} of
 two integer pairs:
 $(i,j)*(j,k)=(i,k)$, where $i,j$ and $k$ are the subscripts, without repetition,
 of the $\lambda$-generators in the commutator and the integer pair has no order,
 {\em i.e.,} $(i,j)=(j,i)$; the multiplication otherwise is either $\{0\}$,
 as two generators commuting, or falls in 
 ${\cal C}$, as two generators
 belonging to one conjugate pair.
 The {\em subscript table} of $su(4)$
 is displayed as,

 \begin{flushleft}
$(1,2)\hspace{0.2cm}(3,4)$\\
$(1,3)\hspace{0.2cm}(2,4)$\\
$(1,4)\hspace{0.2cm}(2,3)$
\end{flushleft}

 \hspace{-0.55cm}which represents the $3$ conjugate pairs of a conjugate partition.
 Simply carrying the conjugate pairs of a conjugate partition,
 the subscript table is orderless in a sense
 that there is no order for integer pairs in one row and
 no order either for these rows in the table.
 Since the multiplication of two integer pairs from any two first rows
 always falls in the rest row, this table carries a quotient algebra too.
 An easy check is that any permutation of the $4$ integers makes the table unchanged,
 noting the orderless of the table. It implies that the binary partitioning
 is invariant {\em w.r.t.} subscript permutations as the dimension $N=4$.

 Now proceed to the subscript table of $su(8)$,

 \begin{flushleft}
 $(1,2) \hspace{0.2cm} (3,4) \hspace{0.2cm} (5,6) \hspace{0.2cm} (7,8) $ \\
 $(1,3) \hspace{0.2cm} (2,4) \hspace{0.2cm} (5,7) \hspace{0.2cm} (6,8) $ \\
 $(1,4) \hspace{0.2cm} (2,3) \hspace{0.2cm} (5,8) \hspace{0.2cm} (6,7) $ \\
 $(1,5) \hspace{0.2cm} (2,6) \hspace{0.2cm} (3,7) \hspace{0.2cm} (4,8) $.
 \end{flushleft}

\hspace{-0.55cm}Only $4$ rows of integer pairs are placed in the table,
 representing the $4$ conjugate pairs.
 For owing to the condition of closure,  the other $3$ rows of integer pairs
 have been decided by these $4$ rows. 
 This is the so-called {\bf \em pre-decision rule} that, for a quotient algebra of $su(2^p)$,
 the $2^p-p-1$ conjugate pairs are determined by certain the other $p$ pairs.
 Prior to the row $\{(1,5)\ (2,6)\ (3,7)\ (4,8)\}$ being concerned,
 any permutations respectively exercised
 in the integer sets $\{1,2,3,4\}$ and $\{5,6,7,8\}$ are allowed.
 For without the consideration of this row,
 these two sets are disjoint and have no ``{\em interaction}."
 While, some inconsistency may be introduced if only a {\em partial} permutation is
 applied to the table with the {\em interaction} row.
 An example is as the table,

 \begin{flushleft}
 $(1,2) \hspace{0.2cm} (3,4) \hspace{0.2cm} (5,7) \hspace{0.2cm} (6,8) $ \\
 $(1,3) \hspace{0.2cm} (2,4) \hspace{0.2cm} (5,6) \hspace{0.2cm} (7,8) $ \\
 $(1,4) \hspace{0.2cm} (2,3) \hspace{0.2cm} (5,8) \hspace{0.2cm} (6,7) $ \\
 $(1,5) \hspace{0.2cm} (2,6) \hspace{0.2cm} (3,7) \hspace{0.2cm} (4,8) $.
 \end{flushleft}

 \hspace{-0.55cm}where the integers $6$ and $7$ are permuted in the first $3$ rows,
 the {\em beginning} rows,  but no corresponding permutation
 is followed in the $4$th row, the {\em interaction} row.
 The order of a row here is referring
 to the display order in the table and has no meaning to the
 conjugate partition.
 This table is literally equivalent to

 \begin{flushleft}
 $(1,2) \hspace{0.2cm} (3,4) \hspace{0.2cm} (5,6) \hspace{0.2cm} (7,8) $ \\
 $(1,3) \hspace{0.2cm} (2,4) \hspace{0.2cm} (5,7) \hspace{0.2cm} (6,8) $ \\
 $(1,4) \hspace{0.2cm} (2,3) \hspace{0.2cm} (5,8) \hspace{0.2cm} (6,7) $ \\
 $(1,5) \hspace{0.2cm} (2,7) \hspace{0.2cm} (3,6) \hspace{0.2cm} (4,8) $.
 \end{flushleft}
 where the original integer pairs are preserved in the first $3$ rows,
 following the binary partitioning, and instead
 integers $6$ and $7$ are permuted in the $4$th row.
 The condition of closure is violated when
 this interaction row is multiplied by either one
 of two beginning rows, the $1$st and the $2$nd rows in the table.
 This example is no atypical. Any permutation of integers in the
 {\em beginning} rows can be rewritten as a permutation
 in the {\em interaction} row and let the integer pairs
 in the beginning rows be obedient to the 
 binary partitioning.
 Then, whether the condition of closure is met is depending on
 if the integer pairs are consistently
 arranged in the interaction row.

 A recursive argument is developed for a general assertion.
 The subscript table of the Lie algebra $su(2^p)$ is prepared by,
 in addition to one interaction row,
 {\em gluing} two such tables of $su(2^{p-1})$ in a way that
 each of the $2^{p-1}-1$ beginning rows is made by
 collecting integer pairs of the identical binary partitioning from
 the two tables. The integer pairs in one table are written in
 integers from $1$ to $2^{p-1}$,
 designated as the $1$st set of integers,
 and the pairs in the other table are in integers from $2^{p-1}+1$ to $2^p$,
 as the $2$nd set.
 The interaction row consists of $2^{p-1}$ integer pairs where
 one entry of each pair is contributed by the $1$st set of integers
 and the other by the $2$nd set.
 Take two integer pairs $(a,b)$ and $(c,d)$ in the interaction row and let
 $a$ and $c$ belong to the $1$st set of integers and $b$ and $d$ to the $2$nd set.
 Suppose that in this row
 at least the two integer pairs $(a,b)$ and $(c,d)$ are inconsistent in
 their associated strings of binary partitioning.
 Two different strings of
 binary partitioning are thus attached to these two integer pairs such that
 $\eta_a+\eta_b\neq \eta_c+\eta_d$, $\eta_a,\dots,\eta_d$ respectively denoting
 $a-1,\dots,d-1$ in their binary expressions. There must be one beginning row that has
 the two integer pairs
 $(b,d)$ and $(\alpha,c)$, leaving the integer $\alpha$ to be decided.
 The two bit-wise additions equal, $\eta_b+\eta_d=\eta_{\alpha}+\eta_c$,
 for these two integer pairs are in one beginning row and
 follow the binary partitioning.
 Then the multiplication of these two rows of integer pairs produces,
 among the others, the pairs $(a,d)$ and $(\alpha,d)$.
 A contradiction occurs,
 $\eta_{\alpha}=\eta_b+\eta_c+\eta_d\neq\eta_a$, that has
 the condition of closure fail.
 It implies that, to keep closed the multiplication of integer pairs,
 the interaction row is obliged to follow the binary partition if
 the beginning rows have been so.
 As known earlier,
 the binary partitioning is naturally respected by $su(4)$, whose
 subscript table is invariant under any integer permutation.
 The recursive argument thus starts with the Lie algebra $su(8), p=3$,
 and has its conjugate pairs constructed according to the binary partitioning.
 Immediately it asserts that, for the Lie algebra $su(2^p)$, only
 the conjugate partition of subscript patterns following the binary partitioning
 allows the structure of the quotient algebra ${\cal \{Q(C};2^p-1)\}$.

 Nevertheless, a {\em global} permutation, applying the same permutation of
 integers to every row of the table, is not denied.
 For in such a case still the binary partitioning is followed
 and integers are permuted simply as labelling indices.
 It hence leads to a conclusion that, up to permutations
 on the subscripts of the $\lambda$-generators, 
 the conjugate partition admitting the structure of the quotient algebra
 ${\cal \{Q(C)\}}$ is unique,
 {\em i.e.}, {\bf\em the sufficient and necessary condition},
 the one whose arrangement of the $\lambda$-generators in its conjugate pairs
 conforms to the binary partitioning.
 Equivalently, the complete set of conjugate partitions that
 accommodate the quotient algebra ${\cal \{Q(C)\}}$
 is obtained by exhaustive permutations of the subscripts of
     $\lambda$-generators initiated by the binary partitioning.
     However, except those coinciding with 
 the binary partitioning,
 the other options have generators written in
 superpositions of noncommuting spinor operators.
 These variations may have advantages in building certain types of Hamiltonians
 for particular physical motivations,
 but are of no interest to the purpose of operator decompositions described later.
 In addition, by an appropriate coordinate transformation,
 the superpositions can be removed and
 the binary partitioning is recovered from these variations.

 \paragraph{Algebra Isomorphism.}
 As the dimension $N$ is not a power of $2$, the Lie algebra $su(N)$
 as well has the structures of the conjugate partition
 and the quotient algebra. The reason is illustrated by an example
 of a conjugate partition for $su(6)$ as given
 in Fig.\ref{Gsu6}.
 Similarly, the partition composed of $7$ conjugate pairs of abelian-subalgebra
 vector spaces and the intrinsic center subalgebra is expressed in the
 $\lambda$-representation.
 The difference between Figs.\ref{Gsu8} and \ref{Gsu6} is that all the
 generators 
 with subscripts larger than the dimension $N=6$,
 {\em i.e.}, with $7$ or $8$ here, are removed from the latter.
 A reminder is that the center subalgebra ${\cal C}$ of $su(8)$ is
 spanned by the set of $7$ independent generators $d_{1l},l=2,\dots,8$.
 The $(i,j)$-th entry of an operator can be referred to as the
 {\em transition-probability amplitude} from the $i$-th to the $j$-th dimensions.
 For $su(6)$, there are no the $7$th and the $8$th dimensions and no transition to these
 two dimensions can be made. Therefore any generators with either one of these
 two subscripts should be set to $0$ or simply removed from the partition.
 As a result, all the rest of $8\times8$ matrices of generators originally for $su(8)$ have only $0$s
 in the $7$th and the $8$th rows and columns.
 Undergoing the removal of the $7$th and the $8$th rows and columns,
 the retaining $35$ independent generators of $6\times6$ matrices suffice for the formation of $su(6)$.
 Conversely, before the $7$th and the $8$th rows and columns are removed,
 these generators support an embedding of $su(6)$ in the space of $su(8)$.

 Since only those with subscripts greater than the dimension $N=6$ are removed,
 still all the $\lambda$-generators with subscripts less than or
 equal to the dimension remain.
 Staying in the original vector spaces, $W_\zeta$ or $\hat{W}_\zeta$,
 of the conjugate pairs or of the center subalgebra ${\cal C}$,
 these retaining generators satisfy
 the condition of closure and the properties of the conjugate partition
 all the same.
 An isomorphic structure of the quotient algebra
 is therefore guaranteed for $su(6)$ and $su(8)$.
 The similar procedure of the {\bf \em removing process} is applicable to
 any Lie algebra $su(2^{p-1}<N<2^p)$. The $\lambda$-generators,
 $\lambda_{ij}$, $\hat{\lambda}_{ij}$ and in
 the set of $2^p-1$ independent $d_{1l},l=2,3,\dots,2^p$, whose either one
 subscript is greater than the dimension $N$ and only they are removed
 from the structure of
 the quotient algebra $\{{\cal Q(C};2^p-1)\}$ given by the intrinsic center subalgebra
 ${\cal C}$ of $su(2^p)$. The Lie algebra $su(N)$ is spanned by
 the remaining $N^2-1$ generators, whose from the $(N+1)$-th to
 the $2^p$-th rows and columns in their matrices are all $0$s and are
 removed. Still all the generators with subscripts ranging from $1$ to $N$
 respectively stay in the conjugate pairs and the center subalgebra
 that are structurally inherited from
 the quotient algebra $\{{\cal Q(C};2^p-1)\}$. 
 The relations among these abelian subalgebras, eqs.\ref{D1}-\ref{D15},
 are accordingly preserved.
 It implies the following Corollary.
 \end{proof}

  For the conjugate partition of the Lie algebra $su(2^{p-1}<N<2^p)$
  obtained by the removing process,
  there are in total $N(N-1)$ generators
  in the conjugate pairs and
  $N-1$ independent generators in the center subalgebra. These two numbers decrease
  compared with those of $su(2^p)$.
  The number, $2^p-1$, of the conjugate pairs and the relations among the pairs
  and the center subalgebra,
  however, remain the same.  The two Lie algebras
  share the identical structure of the quotient algebra.

  \begin{cor}\label{}
  The quotient algebra of the Lie algebra $su(N)$, where $2^{p-1}<N<2^p$
  and $p\in Z^{+}$, is isomorphic to that of $su(2^p)$.
  \end{cor}

\section{Scheme Implementation}
\renewcommand{\theequation}{\arabic{section}.\arabic{equation}}
\setcounter{equation}{0}\hspace{0.3cm}
  The structure of the quotient algebra makes the construction of
  Cartan decompositions straightforward. The essential procedure is to build the table
  of the conjugate partition where the quotient algebra is embedded. With
  the above verification, the algorithm is concluded as follows.

\newpage
 \begin{alg}\label{}
  {\bf Quotient Algebra Construction}
  for $su(N), N\in Z^+${\rm :}\\
  The step of preparation:\\
  For a maximal abelian subalgebra ${\cal A}\in su(N)$, calculate the
  unitary transformation that has ${\cal A}$ diagonalized,
  i.e.,
  $\exists \hspace{0.1cm} U\in SU(N),  \ U{\cal A}U^{\dagger}={\cal C}$,
  mapping ${\cal A}$ to the intrinsic center subalgebra ${\cal C}$.\\
  The 1st step{\rm :}\\
  Take a generator $g_{11}$ outside ${\cal C}$ as a seed and calculate the
  commutator $[g_{11},{\cal C}]$ which produces a finite number, $r_1$,
  of abelian generators
  $\hat{g}_{11},\hat{g}_{12},\cdots,\hat{g}_{1r_1}$, and the vector space
  $\hat{W}_1$ is spanned by this list of generators;
  in the reversing step,
  the commutator $[\hat{g}_{1i},{\cal C}]$, $\hat{g}_{1i}$ being any one generator in
  the list, adds the other $r_1-1$ abelian generators,
  $g_{12},\cdots,g_{1r_1}$, which with the seed form the conjugate space $W_1$. \\
  .........\\
  The $k$-th step{\rm :} \\
  With a generator $g_{k1}$ outside ${\cal C}$
  and the preceding conjugate pairs
  $W_i$ and $\hat{W}_i$, $i=1,2,\dots,k-1$, taken
  as the $k$-th seed,
  calculate the commutator $[g_{k1},{\cal C}]$ that yields the list of abelian generators
  $\hat{g}_{k1},\hat{g}_{k2},\cdots,\hat{g}_{kr_k}$, and the vector space
  $\hat{W}_k$ is spanned by these generators; in the reversing step,
  the commutator $[\hat{g}_{ki},{\cal C}]$,
  $\hat{g}_{ki}$ being any one generator in the list, yields another list of
  $r_k-1$ abelian generators,
  $g_{k2},\cdots,g_{kr_k}$, which with the $k$-th seed form the conjugate space $W_k$.\\
  .........\\
  End when no generator remains.\\
  Merge into one single pair, following the condition of closure,
  any conjugate pairs that commute.\\
  The construction of the quotient algebra $\{{\cal Q(C)}\}$ 
  completes;\\
  Map $\{{\cal Q(C)}\}$ to the quotient algebra $\{{\cal Q(A)}\}$ 
  by $\{{\cal Q(A)}\}=U^{\dagger} \{{\cal Q(C)}\} U$.
  \end{alg}

  The step of merging commuting conjugate pairs is necessary.
  For, in some representations, one conjugate pair may be created by parts
  in the preceding steps.
  The simplest example is in constructing quotient algebras for $su(4)$ in
  the $\lambda$-representation. Due to eqs.\ref{A1} and \ref{A2}, only a single generator,
  its conjugate, is produced by the mapping $[g,{\cal C}]$, where $g$ is
  a $\lambda_{ij}$ or a $\hat{\lambda}_{ij}$ and ${\cal C}$ is the intrinsic
  center subalgebra of $su(4)$. The two options to merge,
  for instance, the two commuting conjugate pairs
  $\{\lambda_{12};\hat{\lambda}_{12}\}$ and $\{\lambda_{34};\hat{\lambda}_{34}\}$
  are $\{\lambda_{12},\lambda_{34};\hat{\lambda}_{12},\hat{\lambda}_{34}\}$
  and $\{\lambda_{12},\hat{\lambda}_{34};\hat{\lambda}_{12},\lambda_{34}\}$.
  Consequently, in addition to the original as in Figs.\ref{csu4} and \ref{Gsu4},
  the $2$nd version
  of the quotient algebra is made as given in Fig.\ref{AGsu4}. This is simply
  the {\em freedom within conjugate pairs},
  $2$ options of distributing each pair of generators
  $\lambda_{ij}$ and $\hat{\lambda}_{ij}$ to their conjugate pair of vector spaces,
  and is loyal to the rule of binary partitioning.
  The generators in the $2$nd version are in a superposition
  of $2$ noncommuting spinor generators, such as
  the part $(\sigma_1\pm \sigma_2)$ for the $2$nd qubit in
  $\lambda_{12}\pm \hat{\lambda}_{34}=\frac{1}{2}(I-\sigma_3)\otimes (\sigma_1\pm \sigma_2)$.
  This form is unfit for the later purpose of operator decompositions,
  although the freedom may be useful in some physical settings.

  In practice, there is no order for the generation of conjugate pairs.
  It is essential and helpful during the construction
  to check the commutation relations
  among the produced pairs by the condition of closure, eq.\ref{D1}.
  At certain stages, for instance, 
  the seed generators needed are easily tracked by applying
  this condition to the existing pairs.

  For the Lie algebra $su(2^p)$, the quotient algebra $\{{\cal Q(A)}\}$
  can be directly constructed with the algorithm
  by taking a maximal abelian subalgebra ${\cal A}$
  as the center subalgebra,
  not necessarily the intrinsic one ${\cal C}$,
  and the number of abelian generators $r_k$ in the
  vector space $W_k$ or $\hat{W}_k$ always equals to $2^{p-1}$.
  Considering the quotient algebra
  given by the intrinsic center subalgebra is particularly
  for the Lie algebra whose dimension is not a power of $2$.
  It is worth a few lines for an equivalence of quotient-algebra construction
  making use of the removing process.
  This version may have the potential to be more efficient when
  the dimension is large.
  Since sharing the isomorphic structure, the quotient algebra $\{{\cal Q(A')}\}$
  given by a maximal abelian subalgebra ${\cal A'}\subset su(N')$
  should be derivable from $\{{\cal Q(C)}\}$, where $N/2<N'<N=2^p$ and ${\cal C}$ is
  the intrinsic center subalgebra of $su(N)$.
  Although requiring two additional steps, the procedure is as simple as follows.
  The first step is to construct $\{{\cal Q(C')}\}$ from $\{{\cal Q(C)}\}$,
  where ${\cal C'}$ is the intrinsic maximal abelian subalgebra of $su(N')$.
  That is to cast $\{{\cal Q(C)}\}$ into the $\lambda$-representation, remove
  all the generators $\lambda_{ij}$ whose $i>N'$ or $j>N'$, and then
  delete all the $(N'+1)$-th through the $N$-th rows and columns of the remaining
  $\lambda$-generators.
  The 2nd step is to map the quotient algebra $\{{\cal Q(C')}\}$
  by a unitary transformation $U'\in SU(N')$ to the coordinate of ${\cal A'}$,
  {\em i.e.}, $\{{\cal Q(A')}\}=U{'}^{\dagger} \{{\cal Q(C')}\} U'$ with
  $U'{\cal A'}{U'}^{\dagger}={\cal C'}$, and finally rewrite the quotient algebra
  $\{{\cal Q(A')}\}$ in the required representation.
  There is an alternative for the transformation.
  Owing to the algebraic isomorphism,
  $\{{\cal Q(A')}\}$ and $\{{\cal Q(A)}\}$ practically share the same transformation $U\in SU(N)$,
  where $U{\cal A'}{U}^{\dagger}={\cal C'}$ and $U{\cal A}{U}^{\dagger}={\cal C}$.
  Here ${\cal A}$ is a unique maximal abelian subalgebra in $su(N)$
  such that ${\cal A'}\subset {\cal A}$, which can be built from ${\cal A'}$
  in terms of the $\lambda$-representation.
  Note that at this point still the quotient algebra $\{{\cal Q(A')}\}$ and
  the intrinsic center subalgebra ${\cal C'}$ are written in $N\times N$ matrices
  of the $\lambda$-generators. The final form of the quotient algebra $\{{\cal Q(A')}\}$
  is attained once the mapping
  $\{{\cal Q(A')}\}=U^{\dagger} \{{\cal Q(C')}\} U$, followed by the deletion of
  all the $(N'+1)$-th through the $N$-th rows and columns,
  and the return to the required representation are completed.

 \paragraph{Cartan Decomposition.}
 Therefore, for any maximal abelian subalgebra
 ${\cal A}\subset su(2^{p-1}<N\leq 2^p)$
 it is feasible to construct a quotient algebra
 $\{{\cal Q(A};2^p-1)\}$
 that has $2^p-1$ conjugate pairs of abelian-subalgebra vector spaces.
 This algebraic structure makes straightforward and systematic
 the construction of the Cartan decomposition.
 The Cartan decomposition
 of the Lie algebra $su(N)$ is referred to
 as the decomposition $su(N)=\mathfrak{t}\oplus \mathfrak{p}$, where
 the subalgebra $\mathfrak{t}$ and the subset $\mathfrak{p}$ satisfy the conditions,
 $[\mathfrak{t},\mathfrak{t}]\subset \mathfrak{t}$,
 $[\mathfrak{t},\mathfrak{p}]\subset \mathfrak{p}$,
 $[\mathfrak{p},\mathfrak{p}]\subset \mathfrak{t}$,
 and ${\rm Tr}\{\mathfrak{t} \mathfrak{p}\}=0$.
 The decomposition is not unique. For a quotient algebra with $2^p-1$ conjugate pairs,
 there are in total $2^p$ choices of
 Cartan decompositions.

  To decide a choice of Cartan decomposition is equivalent to a decision of
  the subalgebra $\mathfrak{t}$.
  Take a quotient algebra of either $su(8)$ or $su(6)$ for example, which has $7$ conjugate
  pairs of vector spaces as shown in Figs.\ref{csu8}-\ref{ncsu6}.
  The subalgebra $\mathfrak{t}$ must be formed by taking either one vector space of
  every conjugate pair.
  However, due to the condition of closure, when the
  first $2$ vector spaces are chosen from the first $2$ conjugate pairs,
  the  $3$rd one from the $3$rd conjugate pair is decided.
  Renote that the order of the conjugate pairs has no relevance. Any $2$ conjugate pairs
  can be specified as the first $2$ pairs, by which the $3$rd pair is determined due to
  the condition of closure.
  The $4$th conjugate pair is arbitrarily specified
  from the remaining pairs, which determines
  the following $5$th through $7$th pairs.
  The vector spaces chosen from the $5$th through the $7$th conjugate pairs are decided
  once the choice from the $4$th conjugate pair is taken.
  For instance, if the set $\{W_1,W_2,\hat{W}_3,\hat{W}_4\}$ is collected,
  the following set of choice is only $\{W_5,W_6,\hat{W}_7\}$.
  The subalgebra $\mathfrak{t}$ is formed by gathering these two sets,
  $\mathfrak{t} = \{W_1,W_2,\hat{W}_3,\hat{W}_4,W_5,W_6,\hat{W}_7\}$.
  The subset $\mathfrak{p}$ is obtained by taking the
  union of the center algebra ${\cal A}$ and the remaining vector space
  in every conjugate pair;
  among the other $7$ ones, ${\cal A}$ is a maximal abelian subalgebra
  in $\mathfrak{p}$.
  Inheriting the properties of spinor generators,
  the subalgebra $\mathfrak{t}$ and the subset $\mathfrak{p}$ fulfill
  the condition of orthogonality,
  ${\rm Tr}\{\mathfrak{t} \mathfrak{p}\}=0$.
  There are in total $2^3$ choices of $\mathfrak{t}$
  for this given quotient algebra.
  In general for a quotient algebra of $2^p-1$ conjugate pairs,
  the subalgebra $\mathfrak{t}$ is decided by the selections of vector spaces,
  a $W_\zeta$ or a $\hat{W}_\zeta$, in the
  $2^r$-th specified conjugate pairs, $r=0,1,2,\dots,p-1$, and thus has in total $2^p$ choices.
  The {\bf \em pre-decision rule} of this form is a result of
  the condition of closure, eq.\ref{D1}.
  Although there are $2^p$ choices of decompositions for a
  quotient algebra of $su(2^{p-1}<N\leq 2^p)$,
  the Cartan decompositions rendered
  by different designations of center subalgebras may have redundancy. Further studies in
  this regard are continued in \cite{TsaiSu}.

\section{Recursive Decomposition}
\renewcommand{\theequation}{\arabic{section}.\arabic{equation}}
 \setcounter{equation}{0}\hspace{0.3cm}
  Forming the subalgebra $\mathfrak{t}$ is only for the $1$st {\em level} of decomposition
  of $su(N)$ and let it be redenoted as $\mathfrak{t}_{[1]}$.
  By the same token, an arbitrary maximal abelian subalgebra in
  $\mathfrak{t}_{[1]}$ can be designated as the center subalgebra
  of the $2$nd level, denoted as ${\cal A}_{[2]}$, which has $\mathfrak{t}_{[1]}$
  partitioned into $2^{p-1}-1$ conjugate pairs.
  Here the subalgebra ${\cal A}_{[2]}$ is either an abelian subalgebra
  in the conjugate pairs
  of the $1$st level or simply a new one created by another group of abelian generators.
  The quotient algebra of the $2$nd level,
  $\{{\cal Q(A}_{[2]};2^{p-1}-1)\}$, is thus constructed.
  With $2^{p-1}$ choices in total,
  the subalgebra $\mathfrak{t}_{[2]}$ for the $2$nd-level decomposition is
  formed in this quotient algebra by applying the condition of closure.
  Similarly, the quotient algebra $\{{\cal Q(A}_{[k]};2^{p-k+1}-1)\}$ is constructed
  at the $k$-th level of decomposition, where
  forming the subalgebra $\mathfrak{t}_{[k]}$ has $2^{p-k+1}$ choices.
  As this process is recursively
  continued for $p$ times to the final level and the quotient algebra
  $\{{\cal Q(A}_{[p]};1)\}$ is derived,
  the Lie algebra $su(2^{p-1}<N\leq 2^p)$ is fully
  decomposed.
  Directing such a recursive decomposition, a {\bf\em decomposition sequence} $seq_{dec}$
  is defined to be the sequence of designated center subalgebras in the order of
  the recursive level and at last
  the subalgebra of the final level,
  $seq_{dec}=\{{\cal A}_{[k]};k=1,2,\cdots,p,p+1, {\cal A}_{[1]}={\cal A}\
  {\rm and}\ {\cal A}_{[p+1]}=\mathfrak{t}_{[p]}\}$.
  Although the decomposition is carried out level by level, the {\bf size},
  the number of generators, of the center subalgebra at each level remains the same.
  For the center subalgebra at the $k$-th level of decomposition
  ${\cal A}_{[k]}$, $1<k\leq p$, is an abelian subalgebra $W_i$ or $\hat{W}_j$
  of the original quotient algebra $\{{\cal Q(A)};2^{p}-1)\}$
  or a new one of a similar size.
  It is easy to verify in terms of the $\lambda$-representation that
  there always exist an enough number of generators in ${\cal A}$ commuting with
  ${\cal A}_{[k]}$ such that the subalgebra
  ${\cal A}_{[k]}$ recovers its $N-1$ abelian generators,
  ref. Appendix B and \cite{TsaiSu}.

 \paragraph{Operator Factorization.}
  The Lie algebra $su(N)$ is recursively decomposable
  based on the structure of the quotient algebra
  residing at each level. According to the $KAK$ theorem \cite{Helgason,Knapp},
  if an algebra
  $\mathfrak{g}$ admits a Cartan decomposition $\mathfrak{g}=\mathfrak{t}\oplus\mathfrak{p}$,
  its corresponding group action $e^{i\mathfrak{g}}$ has a particular form of decomposition.
  That is, $\forall\hspace{0.07cm} {\bf g}\in\mathfrak{g}$,
  $\exists\hspace{0.08cm} {\bf s_1}, {\bf s_2}\in\mathfrak{t}$
  and ${\bf a}\in{\cal A}$, such that
  $e^{i{\bf g}}=e^{i{\bf s_1}} e^{i{\bf a}} e^{i{\bf s_2}}$,
  where ${\cal A}$ is a maximal abelian subalgebra in $\mathfrak{p}$.
  The computation of this decomposition is operationally
  equivalent to the well-known SVD (Singular Value Decomposition) in matrix analysis.
  The essential step is to map the group action to the coordinate
  that has the subalgebra ${\cal A}$ diagonalized. 
  Namely, $\exists\hspace{0.1cm} U\in SU(N)$, s.t.
  $Ue^{i{\bf g}}U^{\dagger}=
  (Ue^{i{\bf s_1}}U^{\dagger})(Ue^{i{\bf a}}U^{\dagger})(Ue^{i{\bf s_2}}U^{\dagger})
   = O_1 D O_2\equiv M$, where $D$ is diagonal and $O_1$ and $O_2\in SO(N)$.
  The matrix $O_1$ is obtained by calculating the diagonalization matrix for
  $M^t=O_1 D^2 {O_1}^t$, and
  $O_2$ obtained by calculating that for $M^t M={O_2}^t D^2 O_2$.
  The entries, the eigenvalues, of $D$ are $e^{i\lambda_j}$ with
  ${\lambda_j}$ being the eigenvalues of ${\bf a}$.
  Such a computation of the group action decomposition,
  or the matrix factorization, is achievable at each level,
  where Cartan decompositions exist owing to the corresponding quotient algebra.
  It implies that the matrix is finally factorized into a product of actions
  contributed by the subalgebras designated in the decomposition sequence.

  According to the factorization guided by the $KAK$ theorem,
  the contributed actions are ordered in the form of a {\em binary bifurcation tree}.
  Let the order of an action in a factorization of
  a unitary transformation $e^{i{\bf g}}\in SU(2^{p-1}<N\leq 2^p)$ be written
  in a $(p+1)$-digit binary string. 
  The contribution of this order can be told
  by the first digit position in the string, traced from the $(p+1)$-th digit,
  where the digit $1$ appears.
  If the final, the $(p+1)$-th, digit is $1$,
  the action of this order is an $e^{i{\bf t}_{[p],r}}, r=1,2,\dots,2^p$, contributed by
  the subalgebra of the final level, ${\bf t}_{[p],r}\in \mathfrak{t}_{[p]}$.
  The action is an $e^{i{\bf a}_{[k],j}}, j=1,2,\dots,2^{k-1}$ and $1\leq k\leq p$,
  contributed by the center subalgebra designated at the $k$-th level,
  ${\bf a}_{[k],j}\in {\cal A}_{[k]}$,
  if $1$ starts at the $k$-th digit.

  {\em Following a decomposition sequence, the recursive decomposition of
  a unitary transformation in $SU(N)$
  implies a {\bf \em path} on the manifold of the group.}
  A subalgebra designated in the sequence may be regarded as
  a {\bf \em post} or a {\bf \em node} during the course of the transformation.
  Interestingly, these posts are visited in a {\bf \em hierarchically recursive order},
  patterned like a bifurcation tree, rather than in a serial order.
  It exhibits the nature of recursive decompositions of Lie groups. 
  Similar to the isomorphism of the quotient algebra, the Lie groups
  $SU(2^{p-1}<N<2^p)$ and $SU(2^p)$ share
  the common spectrum of recursive decompositions,
  or {\em the same chart of decomposition sequences} on the manifold, ref. Appendix B.
  The paths on the manifold are adjustable by changing designations of subalgebras in the
  decomposition sequence or even the group representation up to physical considerations.

  Since every subalgebra in a decomposition sequence is abelian, a unitary transformation
  in $SU(N)$ is fully decomposed into a product of actions $e^{i\omega_\alpha g_\alpha}$,
  where $g_\alpha$s are generators in the chosen representation and $\omega_\alpha$s
  are parameters acquired during the operation of SVD.
  The dimension $N$ has the prime factorization,
  $N=2^{r_0}{P_1}^{r_1}\cdots {P_f}^{r_f}$
  and $P_j, j=1,\dots, f$, being primes greater than $2$.
  The generators of $su(N)$ can be set as tensor products of
  the spinor generators and the generators of $su(P_j)$; here
  the generators of the Lie algebra $su(P_j)$ are expressed in
  the $\lambda$-representation, ref. Appendix A.
  Each contributed action $e^{i\omega_\alpha g_\alpha}$
  plays the role of a quantum gate,
  whose algebra $g_\alpha$ is written
  as a tensor product of generators of $su(2)$ and $su(P_j)$ or
  identity operators of appropriate dimensions.
  For the interest of quantum information,
  a quantum gate $e^{i\omega_\alpha g_\alpha}$ is considered {\em local} if it has only one single
  non-identity operator, {\em i.e.}, a generator of $su(2)$ or $su(P_j)$, in the
  tensor-product form of $g_\alpha$, and considered {\em nonlocal} if otherwise.
  As an immediate result of the decomposition scheme applied to a unitary transformation,
  a theorem of primitive decompositions is concluded as
  follows.

  \paragraph{}
  \hspace{-0.35cm}{\bf Theorem of Gate Decomposition}:
  {\em Every unitary transformation can be fully decomposed into
  a product of local and nonlocal gates.}

  \paragraph{}
  The {\em local-and-nonlocal} decomposition of a transformation is only one application of the scheme.
  More refined results and other applications will be seen in \cite{TsaiSu}.
  This scheme of the Cartan Decomposition enables the
  construction of feasible paths on the manifold of the Lie group $SU(N)$
  with components in a physically required representation.
  The scheme should be of help to
  the exploration in a quantum system.

\bibliographystyle{apalike}

\section*{Appendix}

\appendix
\section{$\lambda$-Representation}
\renewcommand{\theequation}{\Alph{section}.\arabic{equation}}
\setcounter{equation}{0} \hspace{0.5cm}
 A $\lambda$-generator, $\lambda_{ij}$
 or its conjugate $\hat{\lambda}_{ij}$,
 of the $\lambda$-representation is
 an off-diagonal $N\times N$ matrix
 and serves the role of $\sigma_1$ or $\sigma_2$ in
 the $i$-th and the $j$-th dimensions.
 In terms of the Dirac notation, the generators read as
 $\lambda_{ij}=\ket{i}\bra{j}+\ket{j}\bra{i}$ and
 $\hat{\lambda}_{ij}=-i\ket{i}\bra{j}+i\ket{j}\bra{i}$.
 In addition, a diagonal $N\times N$ matrix $d_{kl}$ is
 defined as $d_{kl}=\ket{k}\bra{k}-\ket{l}\bra{l}$.
 For the Lie algebra $su(N)$, a complete set of generators can
 be formed by $N(N-1)/2$ conjugate pairs of
 $\lambda_{ij}$ and $\hat{\lambda}_{ij}$ with
 any $N-1$ independent diagonal operators $d_{kl}$,
 such as $d_{1l}, l=2,3,\dots,N$.
 It is easy to confirm that
 ${\rm Tr}\{\breve{\lambda}_{ij}\breve{\lambda}_{kl}\}=0$ and
 ${\rm Tr}\{\breve{\lambda}_{ij}{\cal C}\}=0$,
 $\breve{\lambda}_{ij}\neq\breve{\lambda}_{kl}$.
 Here $\breve{\lambda}_{ij}$ denotes either
 a $\lambda_{ij}$ or a $\hat{\lambda}_{ij}$,
 and ${\cal C}$ is the intrinsic center subalgebra of $su(N)$.
 The condition of orthogonality ${\rm Tr}\{\mathfrak{t} \mathfrak{p}\}=0$
 is therefore guaranteed at each level of the decomposition,
 for ${\cal C}\subset\mathfrak{p}$.
 Although no such an obligation required in the decomposition,
 the orthogonality can be
 further established in ${\cal C}$ by taking a certain set of basis generators,
 ${\cal C}=span\{\sqrt{\frac{2}{l(l-1)}}\sum_{i=1}^{l-1} d_{il};l=2,3,\dots,N\}$.

 In particular as the dimension $N$ is a prime, the $\lambda$-representation
 is a suitable choice.
 An example are the Gell-Mann matrices for $su(3)$,
 here denoted by $\mu_{j}$, $1\leq j\leq 8$,
\begin{align*}
\mu_{1}=\lambda_{12} \text{,}&
\hspace{0.5cm}\mu_{2}=\hat{\lambda}_{12},\\ \mu_{4}=\lambda_{13}
\text{,}& \hspace{0.5cm}\mu_{5}=\hat{\lambda}_{13},\\
\mu_{6}=\lambda_{23} \text{,}&
\hspace{0.5cm}\mu_{7}=\hat{\lambda}_{23},\\ \mu_{3}=d_{12}
\text{,}& \hspace{0.5cm}\mu_{8}=\frac{1}{\sqrt{3}}(d_{13}+d_{23}).
\end{align*}
 This representation has advantages as well
 when the dimension $N$ admits the prime factorization,
 $N=2^{r_0}{P_1}^{r_1}\cdots {P_f}^{r_f}$
  and $P_j, j=1,\dots, f$, being primes greater than $2$.
 The table of a conjugate partition for $su(2^{r_0})$ can be quickly built
 by taking an arbitrary center subalgebra in the spinor representation.
 An efficient way to build a similar table for
 $su(N)$ is then to make the center subalgebra written in tensor products
 of the spinor generators and the generators of $su(P_j)$ in
 the $\lambda$-representation.

 The following commutation relations characterize the $\lambda$-generators and
 are essential to the construction of conjugate partitions and quotient algebras,
 $1\leq i,j,k,l\leq N$,
\begin{align}
 [\lambda_{ij},d_{kl}]=&\hspace{0.5cm} i\hat{\lambda}_{ij} (-\delta_{ik}+\delta_{il}+\delta_{jk}-\delta_{jl})\label{A1}\\
 [\hat{\lambda}_{ij},d_{kl}]=&\hspace{0.5cm} i\lambda_{ij} (\ \ \delta_{ik}-\delta_{il}-\delta_{jk}+\delta_{jl})\label{A2}\\
 [\lambda_{ij},\hat{\lambda}_{ij}]=&\hspace{0.5cm} 2i d_{ij}\label{A3}\\
[\lambda_{ij},\lambda_{kl}]=&\hspace{0.5cm}
i\hat{\lambda}_{ik}\delta_{jl}+i\hat{\lambda}_{il}\delta_{jk}+i\hat{\lambda}_{jk}\delta_{il}+i\hat{\lambda}_{jl}\delta_{ik}\label{A4}\\
[\lambda_{ij},\hat{\lambda}_{kl}]=&\hspace{0.5cm}
i\lambda_{ik}\delta_{jl}-i\lambda_{il}\delta_{jk}+i\lambda_{jk}\delta_{il}-i\lambda_{jl}\delta_{ik}\label{A5}\\
[\hat{\lambda}_{ij},\hat{\lambda}_{kl}]=&\hspace{0.5cm}
i\hat{\lambda}_{ik}\delta_{jl}-i\hat{\lambda}_{il}\delta_{jk}-i\hat{\lambda}_{jk}\delta_{il}+i\hat{\lambda}_{jl}\delta_{ik}\label{A6}
\end{align}
 For the completion of definition, these generators are either symmetric or antisymmetric
 with respect to the subscript permutation.
 Namely, $\lambda_{ij}=\lambda_{ji}$,
 $\hat{\lambda}_{kl}=-\hat{\lambda}_{lk}$ and $d_{kl}=-d_{lk}$,
 although only those of
 the latter subscript greater than the former are in actual use.

\section{Abelian Subalgebra Extension}
\renewcommand{\theequation}{Alph{section}.\arabic{equation}}
\setcounter{equation}{0}\hspace{0.5cm}
 As aforementioned, a {\em path} on the group manifold is depending on
 designations of subalgebras in the decomposition sequence.
 It is useful to
 search the maximal abelian subalgebras for a Lie algebra.
 Toward this purpose, an easy method based on quotient algebras is described.
 This method considers only the ``{\em basis generators}" and
 ignores those in superpositions of noncommuting operators.
 The intrinsic center subalgebra ${\cal C}$ is
 always a convenient candidate to start with.
 As an example for $su(4)$ shown in Fig.\ref{csu4},
 the center subalgebra
 ${\cal C}=\{\sigma_3\otimes I,I\otimes\sigma_3,\sigma_3\otimes\sigma_3\}$ is laid
 in the central column of the quotient algebra. In this quotient algebra,
 each abelian subalgebra of the conjugate pairs can find $1$ corresponding
 generator in ${\cal C}$ such that the abelian subalgebra recovers to be maximal.
 For instance the maximal abelian subalgebra
 $\{I\otimes\sigma_1,\sigma_3\otimes\sigma_1,\sigma_3\otimes I\}$ is formed by
 taking those in $W_1$ with the generator $\sigma_3\otimes I\in {\cal C}$.
 Immediately, $6$ maximal abelian subalgebras
 of {\em nearest neighbors} are {\em extended} from ${\cal C}$.
 They are reckoned as the maximal abelian subalgebras of
 {\em extension in the $1$st shell}.

 Likewise, the maximal abelian subalgebras
 of the $2$nd-shell extension are constructed from the $6$ quotient algebras given
 by the maximal abelian subalgebras obtained in the $1$st shell.
 However, only $8$ new subalgebras appear
 in this shell and no more new member in the following shells.
 In total there exist $15$ choices of center subalgebras for $su(4)$,
 ignoring those in linear combinations of noncommuting operators.
 Owing to the structure of
 quotient algebras, all the maximal abelian subalgebras
 for the Lie algebra $su(2^{p-1}<N\leq 2^p)$
 can be found in the first $p$ shells \cite{TsaiSu}.
 As the dimension of the Lie algebra is a power of $2$,
 it makes no difference with which maximal abelian subalgebra
 the extension begins.
 While to trace
 those for the Lie algebra $su(2^{p-1}<N<2^p)$, alike to the above algorithm,
 the extension should
 first be performed on the maximal abelian subalgebras of $su(2^p)$ for $p$ shells.
 The version for $su(N)$ emerges after applying
 the removing process to these subalgebras in the $\lambda$-representation
 and then returning the representation required.

 Since each of their maximal abelian subalgebras has $2(2^p-1)$ 
 nearest neighbors and any one of these subalgebras is reachable within
 $p$ shells of extension regardless of the starting member,
 similar to the isomorphism of the quotient algebra,
 the Lie algebras $su(2^p)$ and every $su(2^{p-1}<N<2^p)$ share the identical
 {\em hypercubic structure} of maximal abelian subalgebras.
 The navigation on the manifolds of all the Lie groups $SU(2^{p-1}<N\leq 2^p)$ is hence
 led by only one chart of decomposition sequences.
 More details regarding designations of subalgebras in the sequence and resulted paths
 are given in \cite{TsaiSu}.
 A visionary analogy for Cartan subalgebras may be appropriate and helpful.

\begin{flushleft}
 {\em  Cartan stars, a hypergeometric galaxy serenely shining in the heaven algebra,\\
   shedding light on the unfailing lanes navigating in the group of a baffling sea}
\end{flushleft}

\newpage
\begin{figure}
\begin{center}
\[\begin{array}{ccc}
 \hspace{-0.65cm}(a)\\
\\
I\otimes I\otimes \sigma_{1} & \sigma_{3}\otimes I\otimes I &
I\otimes I\otimes \sigma_{2} \\

&  I \otimes \sigma_{3}\otimes I  &   \sigma_{3} \otimes I\otimes
\sigma_{2}  \\

& I\otimes I\otimes \sigma_{3}  &   I\otimes \sigma_{3}\otimes
\sigma_{2}   \\

& \sigma_{3}\otimes \sigma_{3}\otimes I  &   \sigma_{3}\otimes
\sigma_{3}\otimes \sigma_{2}  \\

& \sigma_{3}\otimes I\otimes \sigma_{3} &  \\

& I\otimes \sigma_{3}\otimes \sigma_{3} & \\

& \sigma_{3}\otimes \sigma_{3}\otimes \sigma_{3} & \\

& & \\
\\
 \hspace{-0.75cm}(b)\\
\\
 I\otimes I\otimes \sigma_{1} &  \sigma_{3}\otimes I\otimes I  & I\otimes I\otimes \sigma_{2} \\

 \sigma_{3}\otimes I\otimes \sigma_{1}  &  I \otimes \sigma_{3}\otimes I  &   \sigma_{3} \otimes I\otimes \sigma_{2}  \\

 I\otimes \sigma_{3}\otimes \sigma_{1} &  I\otimes I\otimes
\sigma_{3} & I\otimes \sigma_{3}\otimes \sigma_{2}    \\

  \sigma_{3}\otimes \sigma_{3}\otimes \sigma_{1} & \sigma_{3}\otimes \sigma_{3}\otimes I  & \sigma_{3}\otimes \sigma_{3}\otimes \sigma_{2}\\

  & \sigma_{3}\otimes I\otimes \sigma_{3} &   \\
& I\otimes \sigma_{3}\otimes \sigma_{3} & \\ & \sigma_{3}\otimes
\sigma_{3}\otimes \sigma_{3} &
\end{array}\]
\end{center}
\caption{(a) With a center subalgebra for $su(8)$ listed in the
central column,
 an abelian subalgebra at the RHS is produced by taking a seed at the LHS.
 (b) In the reversing step, the conjugate abelian subalgebra at the RHS is made by
  taking a seed at the RHS. }\label{stepsu8}
\end{figure}

\newpage
\begin{figure}
\begin{center}
\[\begin{array}{ccccccc}
& & & \cal{C} & & & \\ & & &  \sigma_{3}\otimes I\otimes I  & & &
\\ & & &  I \otimes \sigma_{3}\otimes I  & & &  \\ & & &  I\otimes
I\otimes \sigma_{3}  & & & \\ & & & \sigma_{3}\otimes
\sigma_{3}\otimes I & & & \\ & & & \sigma_{3}\otimes I\otimes
\sigma_{3} & & & \\ & & & I\otimes \sigma_{3}\otimes \sigma_{3} &
& & \\ & & & \sigma_{3}\otimes \sigma_{3}\otimes \sigma_{3} & & &
\\ & W_{1} & & & & \hat{W}_{1} & \\
 I\otimes I\otimes \sigma_{1} & &  \sigma_{3}\otimes I\otimes \sigma_{1}& &  I\otimes I\otimes \sigma_{2}  & &  \sigma_{3}\otimes I\otimes \sigma_{2}\\
 I\otimes \sigma_{3}\otimes \sigma_{1} & &  \sigma_{3}\otimes \sigma_{3}\otimes \sigma_{1} & &  I\otimes \sigma_{3}\otimes \sigma_{2} & &  \sigma_{3}\otimes \sigma_{3}\otimes \sigma_{2} \\
& W_{2} & & & & \hat{W}_{2} & \\
 I\otimes \sigma_{1}\otimes I  & &  \sigma_{3}\otimes \sigma_{1}\otimes I  & &  I\otimes \sigma_{2}\otimes I  &&  \sigma_{3}\otimes \sigma_{2}\otimes I  \\
 I\otimes \sigma_{1}\otimes \sigma_{3}  & &  \sigma_{3}\otimes \sigma_{1}\otimes \sigma_{3}  & &  I\otimes \sigma_{2}\otimes \sigma_{3}  & &  \sigma_{3}\otimes \sigma_{2}\otimes \sigma_{3} \\
& W_{3} & & & & \hat{W}_{3} & \\
 I\otimes \sigma_{1}\otimes \sigma_{1}  & &  I\otimes \sigma_{2}\otimes \sigma_{2}  & &  I\otimes \sigma_{2}\otimes \sigma_{1}  & &  I\otimes \sigma_{1}\otimes \sigma_{2} \\
 \sigma_{3}\otimes \sigma_{1}\otimes \sigma_{1}  & &  \sigma_{3}\otimes \sigma_{2}\otimes \sigma_{2}  & &  \sigma_{3}\otimes \sigma_{2}\otimes \sigma_{1}  & &  \sigma_{3}\otimes \sigma_{1}\otimes \sigma_{2} \\
& W_{4} & & & & \hat{W}_{4} & \\
 \sigma_{1}\otimes I\otimes I  & &  \sigma_{1}\otimes \sigma_{3}\otimes I  &  &  \sigma_{2}\otimes I\otimes I  & &  \sigma_{2}\otimes \sigma_{3}\otimes I  \\
 \sigma_{1}\otimes I\otimes \sigma_{3}  & &  \sigma_{1}\otimes \sigma_{3}\otimes \sigma_{3}  & &  \sigma_{2}\otimes I\otimes \sigma_{3}  & &  \sigma_{2}\otimes \sigma_{3}\otimes \sigma_{3}  \\
& W_{5} & & & & \hat{W}_{5} & \\
 \sigma_{1}\otimes I\otimes \sigma_{1}  & &  \sigma_{2}\otimes I\otimes \sigma_{2}  & &  \sigma_{2}\otimes I\otimes \sigma_{1}  & &  \sigma_{1}\otimes I\otimes \sigma_{2} \\
 \sigma_{1}\otimes \sigma_{3}\otimes \sigma_{1}  & &  \sigma_{2}\otimes \sigma_{3}\otimes \sigma_{2}  & &  \sigma_{2}\otimes \sigma_{3}\otimes \sigma_{1}  & &  \sigma_{1}\otimes \sigma_{3}\otimes \sigma_{2} \\
& W_{6} & & & & \hat{W}_{6} & \\
 \sigma_{1}\otimes \sigma_{1}\otimes I  & &  \sigma_{2}\otimes \sigma_{2}\otimes I  & &  \sigma_{2}\otimes \sigma_{1}\otimes I  & &  \sigma_{1}\otimes \sigma_{2}\otimes I \\
 \sigma_{1}\otimes \sigma_{1}\otimes \sigma_{3}  & &  \sigma_{2}\otimes \sigma_{2}\otimes \sigma_{3}  & &  \sigma_{2}\otimes \sigma_{1}\otimes \sigma_{3}  & &  \sigma_{1}\otimes \sigma_{2}\otimes \sigma_{3} \\
& W_{7} & & & & \hat{W}_{7} & \\
 \sigma_{1}\otimes \sigma_{1}\otimes \sigma_{1} & &  \sigma_{2}\otimes \sigma_{2}\otimes \sigma_{1}  & &  \sigma_{2}\otimes \sigma_{1}\otimes \sigma_{1}  & & \sigma_{1}\otimes \sigma_{2}\otimes \sigma_{1} \\
 \sigma_{2}\otimes \sigma_{1}\otimes \sigma_{2}  & &
\sigma_{1}\otimes \sigma_{2}\otimes \sigma_{2} & &
\sigma_{1}\otimes \sigma_{1}\otimes \sigma_{2}  & &
\sigma_{2}\otimes \sigma_{2}\otimes \sigma_{2}
\end{array} \]
\end{center}
\caption{A conjugate partition and a quotient algebra given
 by the intrinsic center subalgebra of $su(8)$.}\label{csu8}
\end{figure}

\newpage
\begin{figure}
\begin{center}
\[\begin{array}{ccccccc}
 & & &  I\otimes \sigma_{3}  & & &\\
 & & &   \mu_{3} \otimes I   & & & \\
 & & &   \mu_{8} \otimes I  & & & \\
 & & & \mu_{3} \otimes \sigma_{3} & & & \\
& & & \mu_{8} \otimes \sigma_{3} & & & \\ & W_{1} & & & &
\hat{W}_{1} & \\
 I \otimes \sigma_{1}  &  \mu_{3}\otimes \sigma_{1}  &  \mu_{8}\otimes \sigma_{1}  & &  I\otimes \sigma_{2}  &  \mu_{3}\otimes \sigma_{2}  &  \mu_{8} \otimes \sigma_{2} \\
& W_{2} & & & & \hat{W}_{2} & \\
  \mu_{1} \otimes I  & &  \mu_{1}\otimes\sigma_{3} & &  \mu_{2} \otimes I  & & \mu_{2} \otimes \sigma_{3}\\
& W_{3} & & & & \hat{W}_{3} & \\
   \mu_{1} \otimes \sigma_{1} & & \mu_{2}\otimes \sigma_{2} & &  \mu_{1}  \otimes \sigma_{2}  & & \mu_{2}\otimes \sigma_{1}\\
& W_{4} & & & & \hat{W}_{4} & \\
  \mu_{4} \otimes I  & & \mu_{4}\otimes \sigma_{3} & &   \mu_{5} \otimes I  & & \mu_{5} \otimes \sigma_{3}\\
& W_{5} & & & & \hat{W}_{5} & \\
  \mu_{4} \otimes \sigma_{1}  & & \mu_{5}\otimes \sigma_{2} & &  \mu_{4} \otimes \sigma_{2}  & & \mu_{5}\otimes \sigma_{1}\\
& W_{6} & & & & \hat{W}_{6} & \\
  \mu_{6} \otimes I  & & \mu_{6}\otimes \sigma_{3} & &   \mu_{7} \otimes I  & & \mu_{7}\otimes \sigma_{3}\\
& W_{7} & & & & \hat{W}_{7} & \\
  \mu_{6} \otimes  \sigma_{1} & & \mu_{7}\otimes \sigma_{2} &
&   \mu_{6}\otimes  \sigma_{2} & & \mu_{7}\otimes \sigma_{1}
\end{array}\]
\caption{A conjugate partition and a quotient algebra given by the
intrinsic center subalgebra of $su(6)$,
 the generators $\mu_j$ denoting the Gell-Mann matrices, and
 the former $I$, before the symbol $\otimes$,
 being the $3\times 3$ identity in contrast to
 the $2\times 2$ identity of the latter.}\label{csu6}
\end{center}
\end{figure}

\newpage
\begin{figure}
\begin{center}
\[\begin{array}{ccccccc}
& & &  \sigma_{1}\otimes I\otimes I  & & & \\ & & &  I \otimes
\sigma_{1}\otimes I  & & &  \\ & & &  I\otimes I\otimes \sigma_{1}
& & & \\ & & & \sigma_{1}\otimes \sigma_{1}\otimes I & & & \\ & &
& \sigma_{1}\otimes I\otimes \sigma_{1} & & & \\ & & & I\otimes
\sigma_{1}\otimes \sigma_{1} & & & \\ & & & \sigma_{1}\otimes
\sigma_{1}\otimes \sigma_{1} & & & \\ & W_{1} & & & & \hat{W}_{1}
& \\
 \sigma_{3}\otimes I\otimes I  & &  \sigma_{3}\otimes \sigma_{1}\otimes I  &  &  \sigma_{2}\otimes I\otimes I  & &  \sigma_{2}\otimes \sigma_{1}\otimes I  \\
 \sigma_{3}\otimes I\otimes \sigma_{1}  & &  \sigma_{3}\otimes \sigma_{1}\otimes \sigma_{1}  & &  \sigma_{2}\otimes I\otimes \sigma_{1}  & &  \sigma_{2}\otimes \sigma_{1}\otimes \sigma_{1}  \\
& W_{2} & & & & \hat{W}_{2} & \\
 I\otimes \sigma_{3}\otimes I  & &  \sigma_{1}\otimes \sigma_{3}\otimes I  & &  I\otimes \sigma_{2}\otimes I  &&  \sigma_{1}\otimes \sigma_{2}\otimes I  \\
 I\otimes \sigma_{3}\otimes \sigma_{1}  & &  \sigma_{1}\otimes \sigma_{3}\otimes \sigma_{1}  & &  I\otimes \sigma_{2}\otimes \sigma_{1}  & &  \sigma_{1}\otimes \sigma_{2}\otimes \sigma_{1} \\
& W_{3} & & & & \hat{W}_{3} & \\
 I\otimes I\otimes \sigma_{3} & &  \sigma_{1}\otimes I\otimes \sigma_{3}& &  I\otimes I\otimes \sigma_{2}  & &  \sigma_{1} \otimes I\otimes \sigma_{2}\\
 I\otimes \sigma_{1}\otimes \sigma_{3} & &  \sigma_{1}\otimes \sigma_{1}\otimes \sigma_{3} & &  I\otimes \sigma_{1}\otimes \sigma_{2} & &  \sigma_{1}\otimes \sigma_{1}\otimes \sigma_{2} \\
 & W_{4} & & & & \hat{W}_{4} & \\
 \sigma_{3}\otimes \sigma_{3}\otimes I  & &  \sigma_{2}\otimes \sigma_{2}\otimes I  & &  \sigma_{2}\otimes \sigma_{3}\otimes I  & &  \sigma_{3}\otimes \sigma_{2}\otimes I \\
 \sigma_{3}\otimes \sigma_{3}\otimes \sigma_{1}  & &  \sigma_{2}\otimes \sigma_{2}\otimes \sigma_{1}  & &  \sigma_{2}\otimes \sigma_{3}\otimes \sigma_{1}  & & \sigma_{3}\otimes \sigma_{2}\otimes \sigma_{1} \\
 & W_{5} & & & & \hat{W}_{5} & \\
 \sigma_{3}\otimes I\otimes \sigma_{3}  & &  \sigma_{2}\otimes I\otimes \sigma_{2}  & &  \sigma_{2}\otimes I\otimes \sigma_{3}  & &  \sigma_{3}\otimes I\otimes \sigma_{2} \\
 \sigma_{3}\otimes \sigma_{1}\otimes \sigma_{3}  & &  \sigma_{2}\otimes \sigma_{1}\otimes \sigma_{2}  & &  \sigma_{2}\otimes \sigma_{1}\otimes \sigma_{3}  & &  \sigma_{3}\otimes \sigma_{1}\otimes \sigma_{2} \\
 & W_{6} & & & & \hat{W}_{6} & \\
 I\otimes \sigma_{3}\otimes \sigma_{3}  & &  I\otimes \sigma_{2}\otimes \sigma_{2}  & &  I\otimes \sigma_{2}\otimes \sigma_{3}  & &  I\otimes \sigma_{3}\otimes \sigma_{2} \\
 \sigma_{1}\otimes \sigma_{3}\otimes \sigma_{3}  & &  \sigma_{1}\otimes \sigma_{2}\otimes \sigma_{2}  & &  \sigma_{1}\otimes \sigma_{2}\otimes \sigma_{3}  & &  \sigma_{1}\otimes \sigma_{3}\otimes \sigma_{2} \\
 & W_{7} & & & & \hat{W}_{7} & \\
 \sigma_{3}\otimes \sigma_{3}\otimes \sigma_{3}  & &  \sigma_{2}\otimes \sigma_{2}\otimes \sigma_{3}  & &  \sigma_{2}\otimes \sigma_{3}\otimes \sigma_{3}  & &  \sigma_{3}\otimes \sigma_{2}\otimes \sigma_{3} \\
 \sigma_{2}\otimes \sigma_{3}\otimes \sigma_{2}  & &
\sigma_{3}\otimes \sigma_{2}\otimes \sigma_{2}  & &
\sigma_{3}\otimes \sigma_{3}\otimes \sigma_{2}  & &
\sigma_{2}\otimes \sigma_{2}\otimes \sigma_{2}
\end{array}\]
\end{center}
\caption{A conjugate partition and a quotient algebra given by
 a non-diagonal center subalgebra of $su(8)$.}\label{ncsu8}
\end{figure}

\newpage

\newpage
\begin{figure}
\begin{center}
\[\begin{array}{ccc}
& diag\{1, 1, 1, 1,-1,-1,-1,-1 \} & \\ & diag\{1, 1,-1,-1, 1,
1,-1,-1 \} &\\ & diag\{1,-1, 1,-1, 1,-1, 1,-1 \} &\\ & diag\{1,
1,-1,-1,-1,-1, 1, 1 \} &\\ & diag\{1,-1, 1,-1,-1, 1,-1, 1 \} &\\ &
diag\{1,-1,-1, 1, 1,-1,-1, 1 \} &\\ & diag\{1,-1,-1, 1,-1, 1, 1,-1
\} &\\ W_{001} & & \hat{W}_{001} \\
 \lambda_{12}+\lambda_{34}+\lambda_{56}+\lambda_{78}  & &   \hat{\lambda}_{12}+\hat{\lambda}_{34}+\hat{\lambda}_{56}+\hat{\lambda}_{78}    \\
\lambda_{12}+\lambda_{34}-\lambda_{56}-\lambda_{78}   & &
\hat{\lambda}_{12}+\hat{\lambda}_{34}-\hat{\lambda}_{56}-\hat{\lambda}_{78}
\\
 \lambda_{12}-\lambda_{34}+\lambda_{56}-\lambda_{78}  & &     \hat{\lambda}_{12}-\hat{\lambda}_{34}+\hat{\lambda}_{56}-\hat{\lambda}_{78}   \\
 \lambda_{12}-\lambda_{34}-\lambda_{56}+\lambda_{78}  & &   \hat{\lambda}_{12}-\hat{\lambda}_{34}-\hat{\lambda}_{56}+\hat{\lambda}_{78}\\
W_{010} & & \hat{W}_{010} \\
 \lambda_{13}+\lambda_{24}+\lambda_{57}+\lambda_{68}  & &   \hat{\lambda}_{13}+\hat{\lambda}_{24}+\hat{\lambda}_{57}+\hat{\lambda}_{68}    \\
\lambda_{13}+\lambda_{24}-\lambda_{57}-\lambda_{68}   & &
\hat{\lambda}_{13}+\hat{\lambda}_{24}-\hat{\lambda}_{57}-\hat{\lambda}_{68}
\\
 \lambda_{13}-\lambda_{24}+\lambda_{57}-\lambda_{68}  & &     \hat{\lambda}_{13}-\hat{\lambda}_{24}+\hat{\lambda}_{57}-\hat{\lambda}_{68}   \\
 \lambda_{13}-\lambda_{24}-\lambda_{57}+\lambda_{68}  & &   \hat{\lambda}_{13}-\hat{\lambda}_{24}-\hat{\lambda}_{57}+\hat{\lambda}_{68}\\
W_{011} & & \hat{W}_{011} \\
 \lambda_{14}+\lambda_{23}+\lambda_{58}+\lambda_{67}  & &   \hat{\lambda}_{14}+\hat{\lambda}_{23}+\hat{\lambda}_{58}+\hat{\lambda}_{67}    \\
-\lambda_{14}+\lambda_{23}-\lambda_{58}+\lambda_{67}   & &
\hat{\lambda}_{14}-\hat{\lambda}_{23}+\hat{\lambda}_{58}-\hat{\lambda}_{67}
\\
 \lambda_{14}+\lambda_{23}-\lambda_{58}-\lambda_{67}  & &     \hat{\lambda}_{14}+\hat{\lambda}_{23}-\hat{\lambda}_{58}-\hat{\lambda}_{67}   \\
 -\lambda_{14}+\lambda_{23}+\lambda_{58}-\lambda_{67}  & &   \hat{\lambda}_{14}-\hat{\lambda}_{23}-\hat{\lambda}_{58}+\hat{\lambda}_{67}\\
W_{100} & & \hat{W}_{100} \\
 \lambda_{15}+\lambda_{26}+\lambda_{37}+\lambda_{48}  & &   \hat{\lambda}_{15}+\hat{\lambda}_{26}+\hat{\lambda}_{37}+\hat{\lambda}_{48}    \\
\lambda_{15}+\lambda_{26}-\lambda_{37}-\lambda_{48}   & &
\hat{\lambda}_{15}+\hat{\lambda}_{26}-\hat{\lambda}_{37}-\hat{\lambda}_{48}
\\
 \lambda_{15}-\lambda_{26}+\lambda_{37}-\lambda_{48}  & &     \hat{\lambda}_{15}-\hat{\lambda}_{26}+\hat{\lambda}_{37}-\hat{\lambda}_{48}   \\
 \lambda_{15}-\lambda_{26}-\lambda_{37}+\lambda_{48}  & &   \hat{\lambda}_{15}-\hat{\lambda}_{26}-\hat{\lambda}_{37}+\hat{\lambda}_{48}\\
W_{101} & & \hat{W}_{101} \\
 \lambda_{16}+\lambda_{25}+\lambda_{38}+\lambda_{47}  & &   \hat{\lambda}_{16}+\hat{\lambda}_{25}+\hat{\lambda}_{38}+\hat{\lambda}_{47}    \\
-\lambda_{16}+\lambda_{25}-\lambda_{38}+\lambda_{47}   & &
\hat{\lambda}_{16}-\hat{\lambda}_{25}+\hat{\lambda}_{38}-\hat{\lambda}_{47}
\\
 \lambda_{16}+\lambda_{25}-\lambda_{38}-\lambda_{47}  & &     \hat{\lambda}_{16}+\hat{\lambda}_{25}-\hat{\lambda}_{38}-\hat{\lambda}_{47}   \\
 -\lambda_{16}+\lambda_{25}+\lambda_{38}-\lambda_{47}  & &   \hat{\lambda}_{16}-\hat{\lambda}_{25}-\hat{\lambda}_{38}+\hat{\lambda}_{47}\\
W_{110} & & \hat{W}_{110} \\
 \lambda_{17}+\lambda_{28}+\lambda_{35}+\lambda_{46}  & &   \hat{\lambda}_{17}+\hat{\lambda}_{28}+\hat{\lambda}_{35}+\hat{\lambda}_{46}    \\
-\lambda_{17}-\lambda_{28}+\lambda_{35}+\lambda_{46}   & &
\hat{\lambda}_{17}+\hat{\lambda}_{28}-\hat{\lambda}_{35}-\hat{\lambda}_{46}
\\
 \lambda_{17}-\lambda_{28}+\lambda_{35}-\lambda_{46}  & &     \hat{\lambda}_{17}-\hat{\lambda}_{28}+\hat{\lambda}_{35}-\hat{\lambda}_{46}   \\
 -\lambda_{17}+\lambda_{28}+\lambda_{35}-\lambda_{46}  & &   \hat{\lambda}_{17}-\hat{\lambda}_{28}-\hat{\lambda}_{35}+\hat{\lambda}_{46}\\
W_{111} & & \hat{W}_{111} \\
 \lambda_{18}+\lambda_{27}+\lambda_{36}+\lambda_{45} & &   \hat{\lambda}_{18}+\hat{\lambda}_{27}+\hat{\lambda}_{36}+\hat{\lambda}_{45}    \\
-\lambda_{18}-\lambda_{27}+\lambda_{36}+\lambda_{45}   & &
\hat{\lambda}_{18}+\hat{\lambda}_{27}-\hat{\lambda}_{36}-\hat{\lambda}_{45}
\\
 -\lambda_{18}+\lambda_{27}-\lambda_{36}+\lambda_{45}  & &     \hat{\lambda}_{18}-\hat{\lambda}_{27}+\hat{\lambda}_{36}-\hat{\lambda}_{45}   \\
 -\lambda_{18}+\lambda_{27}+\lambda_{36}-\lambda_{45}  & &
-\hat{\lambda}_{18}-\hat{\lambda}_{27}+\hat{\lambda}_{36}+\hat{\lambda}_{45}
\end{array} \]
\caption{The conjugate partition of Fig.\protect\ref{csu8}
 in the $\lambda$-representation.}\label{Gsu8}
\end{center}
\end{figure}

\newpage
\begin{figure}
\begin{center}
\[\begin{array}{ccc}
& diag\{1,-1, 1,-1,1,-1 \} & \\ & diag\{1, 1,-1,-1, 0,0 \} &\\ &
diag\{1,1, 1,1, -2,-2 \} &\\ & diag\{1, -1,-1,1,0,0 \}  &\\ &
diag\{1,-1, 1,-1,-2,2 \} &\\ W_{001} &  &\hat{W}_{001} \\
\lambda_{12}-\lambda_{34} & &
\hat{\lambda}_{12}-\hat{\lambda}_{34}  \\
\lambda_{12}+\lambda_{34}+\lambda_{56} &
&\hat{\lambda}_{12}+\hat{\lambda}_{34}+\hat{\lambda}_{56}  \\
\lambda_{12}+\lambda_{34}-2\lambda_{56}&&\hat{\lambda}_{12}+\hat{\lambda}_{34}-2\hat{\lambda}_{56}
\\

W_{010} &  &\hat{W}_{010} \\
 \lambda_{13}+\lambda_{24}&  &  \hat{\lambda}_{13}+\hat{\lambda}_{24}  \\
\lambda_{13}-\lambda_{24} &
&\hat{\lambda}_{13}-\hat{\lambda}_{24}\\ W_{011} & &\hat{W}_{011}
\\
 \lambda_{14}+\lambda_{23} &   &  \hat{\lambda}_{14}-\hat{\lambda}_{23}  \\
-\lambda_{14}+\lambda_{23} &  &
\hat{\lambda}_{14}+\hat{\lambda}_{23}    \\

W_{100} &  &\hat{W}_{100} \\
 \lambda_{15}+\lambda_{26} &  &  \hat{\lambda}_{15}+\hat{\lambda}_{26}  \\
 \lambda_{15}-\lambda_{26} &   &  \hat{\lambda}_{15}-\hat{\lambda}_{26} \\                                                                                     \\

W_{101} &  &\hat{W}_{101} \\
 \lambda_{16}+\lambda_{25} &   & \hat{\lambda}_{16}+\hat{\lambda}_{25} \\
 -\lambda_{16}+\lambda_{25} &   & \hat{\lambda}_{16}-\hat{\lambda}_{25} \\

W_{110} & &\hat{W}_{110} \\ \lambda_{35}+\lambda_{46}  &  &
\hat{\lambda}_{35}+\hat{\lambda}_{46}   \\
\lambda_{35}-\lambda_{46}  &  &
\hat{\lambda}_{35}-\hat{\lambda}_{46}    \\

W_{111} & &\hat{W}_{111} \\ \lambda_{36}+\lambda_{45}  &   &
\hat{\lambda}_{36}+\hat{\lambda}_{45} \\
-\lambda_{36}+\lambda_{45}  &  &
\hat{\lambda}_{36}-\hat{\lambda}_{45}
\end{array}\] \caption{The conjugate partition of Fig.\protect\ref{csu6}
 in the $\lambda$-representation.}\label{Gsu6}
\end{center}
\end{figure}

\newpage

\newpage
\begin{figure}
\begin{center}
\[\begin{array}{ccccccc}
&& & \sigma_{3}\otimes I  &&&\\
 && & I\otimes \sigma_{3}  &&&\\
 && & \sigma_{3}\otimes \sigma_{3} &&&\\
 & W_{1} &&&& \hat{W}_{1}& \\
  I\otimes \sigma_{1} &  & \sigma_{3}\otimes \sigma_{1} && I\otimes \sigma_{2}   & & \sigma_{3}\otimes \sigma_{2}\\
& W_{2} && & &\hat{W}_{2}& \\ \sigma_{1}\otimes I  &  &
\sigma_{1}\otimes \sigma_{3} &&   \sigma_{2}\otimes I  &
&\sigma_{2}\otimes \sigma_{3} \\ & W_{3} && && \hat{W}_{3}& \\
 \sigma_{1}\otimes \sigma_{1} &  & \sigma_{2}\otimes\sigma_{2}  & &  \sigma_{2}\otimes \sigma_{1}  & & \sigma_{1}\otimes \sigma_{2}
\end{array}\]\\
\caption{A conjugate partition and a quotient algebra
 given by the intrinsic center subalgebra of $su(4)$.}\label{csu4}
\end{center}
\end{figure}

\begin{figure}
\begin{center}
\[\begin{array}{ccccccc}
&& & diag\{ 1,1,-1,-1 \}  &&&\\ && & diag\{1,-1,1,-1 \}  &&&\\ &&
& diag\{1,-1,-1,1 \}  &&&\\ & W_{01} &&&& \hat{W}_{01}& \\
\lambda_{12}+\lambda_{34} & & \lambda_{12}-\lambda_{34} & &
\hat{\lambda}_{12}+\hat{\lambda}_{34} &  &
\hat{\lambda}_{12}-\hat{\lambda}_{34}\\ & W_{10} &&&&
\hat{W}_{10}& \\ \lambda_{13}+\lambda_{24} & &
\lambda_{13}-\lambda_{24} & &
\hat{\lambda}_{13}+\hat{\lambda}_{24} &  &
\hat{\lambda}_{13}-\hat{\lambda}_{24}\\ & W_{11} &&&&
\hat{W}_{11}& \\ \lambda_{14}+\lambda_{23} & &
\lambda_{14}-\lambda_{23} & &
\hat{\lambda}_{14}+\hat{\lambda}_{23} &  &
\hat{\lambda}_{14}-\hat{\lambda}_{23}
\end{array} \]
\caption{The partition and algebra of Fig.\protect\ref{csu4} in
the $\lambda$-representation.}\label{Gsu4}
\end{center}
\end{figure}

\begin{figure}
\begin{center}

\[\begin{array}{ccccccc}

\end{array}\]\\

\caption{An alternative of the conjugate partition and the
quotient algebra for that in Fig.\protect\ref{Gsu4}.}\label{AGsu4}
\end{center}
\end{figure}

\end{document}